\documentclass[%
 aip,
 amsmath,amssymb,
 reprint,%
]{revtex4-1}

\usepackage{graphicx}% Include figure files
\usepackage{dcolumn}% Align table columns on decimal point
\usepackage{bm}% bold math
\usepackage[utf8]{inputenc}
\usepackage[T1]{fontenc}
\usepackage{mathptmx}
\usepackage{color}
\usepackage{appendix}

\begin{document}

\preprint{AIP/123-QED}

\title[]{
%Modeling Single Excitations
%with an Excited State Variational Principle
%in Density Functional Theory
A Density Functional Extension to Excited State Mean-Field Theory
%A Variational Formalism for Excited State Density Functional Theory
}
% Force line breaks with \\

\author{Luning Zhao}
\affiliation{
Department of Chemistry, University of California, Berkeley, California 94720, USA 
}
\author{Eric Neuscamman}
 \email{eneuscamman@berkeley.edu.}
\affiliation{
Department of Chemistry, University of California, Berkeley, California 94720, USA 
}
\affiliation{Chemical Sciences Division, Lawrence Berkeley National Laboratory, Berkeley, CA, 94720, USA}

\date{\today}% It is always \today, today,

\begin{abstract}
%{\color{red} Better CT energies than RSH TDDFT even when using non-RSH functionals.}
We investigate an extension of excited state mean-field theory in which the energy expression
is augmented with density functional components in an effort to include the effects of weak
electron correlations.
%We present a density functional extension to the excited state mean-field theory. 
The approach remains variational and entirely time-independent, allowing it to avoid
some of the difficulties associated with linear response and the adiabatic approximation.
In particular, all of the electrons' orbitals are relaxed state specifically
and there is no reliance on Kohn-Sham orbital energy differences,
both of which are important features in the context of charge transfer.
%In this method, a minimally correlated wave function with correct spin and spatial
%symmetry is used to compute kinetic, external potential, and electron-electron
%repulsion energy. The electron correlation energy is separated into weak and 
%strong parts and are treated by density functional and wave function respectively. 
%for a variational excited state
%density functional theory with strong parallels to ground state Kohn-Sham theory.
%In particular, the approach develops density functional expressions
%for the energies of singly-excited states and combines them with
%the variational principle from excited state mean field theory.
%This method is built upon the recently developed 
%excited state variational principle and the excited state mean-field platform.
%{\color{red}
%As in the ground state case, the approach is formally exact but also lends
%itself to practical execution via familiar approximations for the density
%functional. (Not sure this is true)
%}
%This formalism avoids the need for linear response and the adiabatic
%approximation and thus
%addresses both the electron-affinity/ionization-potential imbalance and the need for post-excitation orbital relaxations.
Preliminary testing shows clear advantages for single-component charge
transfer states, but the method, at least in its current form, is less reliable
for states in which multiple particle-hole transitions contribute
significantly.
%these features deliver dramatic
%improvements in charge transfer energies relative to time dependent density
%functional theory while maintaining accuracy for valence excitations.
%which we show lead to
%accuracies in charge transfer excitations that far exceed those of even
%a modern range-separated approach to time
% Additional correlation energy that arises due to the multi-reference nature
%of the fictitious system is manually added to account for spin symmetry and static 
%correlation effect. By allowing the charge density of the excited states to relax according to 
%the excitation, we 
%are able to go beyond the adiabatic approximation made by the time-dependent density functional
%theory (TDDFT). In charge-transfer excited states where the adiabatic approximation produces substantial
%amount of errors, our method yields accurate predictions to both the excitation energy values and 
%their asymptotic behaviors, surpassing the accuracy of the range-separated hybrid functionals. 
\end{abstract}

\maketitle

\section{Introduction}
\label{sec:intro}

The recently developed excited state mean-field theory (ESMF)\cite{Shea2018} 
is intended to act as a mean-field platform for excited states in much the same
way as  Hartree-Fock (HF) theory \cite{Szabo-Ostland}
does for ground states.
As one might expect, these two theories share many properties:
they rely on minimally correlated wave function forms,
produce energy stationary points, relax orbital shapes variationally,
and have the same cost-scaling.
Also like HF theory, ESMF lacks a proper description of correlation effects,
and so from a practical standpoint is expected to be more useful as a starting point for
correlation methods than as a way of making energy predictions on its own.
While there are many ways one could go about capturing correlation effects,
it is hard to avoid thinking about density functional theory (DFT) \cite{Parr-Yang}
in this context given how useful it is for this purpose in ground states.

The Kohn-Sham (KS) formulation of DFT \cite{ks-dft} is the most widely used 
electronic structure method in chemistry, physics, and materials science. Due to its
favorable scaling with system size and reasonable accuracy in a variety of different
circumstances, DFT is often regarded as one of the most powerful tools for studying 
the electronic and dynamic properties of materials and medium to large molecules.
The KS-DFT method can also be considered as an extension to the HF method, by 
replacing the exchange energy in HF with the exchange-correlation (xc) energy in
KS-DFT. With the exact xc functional, KS-DFT is able to capture correlation
effects exactly. Comparing to other post-HF methods that account for weak
correlation effects, such as configuration interaction, Moller-Plesset 2nd order 
perturbation theory, and coupled cluster, the most appealing feature of KS-DFT is
its low cost-scaling, which allows it to be applied to systems with thousands of 
electrons or more. 

Inspired by the success of KS-DFT in ground states, one may wonder whether similar
extensions using DFT can also be achieved for ESMF. Intuitively, combining ESMF with
DFT would allow one to go beyond the mean-field form of the ESMF wave function
and be able to recover weak correlation effects while maintaining the mean-field
cost scaling of ESMF. More importantly, such an approach need not rely on linear 
response (LR) theory or the adiabatic approximation (AA), both of which are
central to the practical application of time-dependent density functional theory
(TDDFT)\cite{Runge1984,casida2012,Ullrich2013,Burke2004}.
As the combination of LR and the AA can produce significant errors in some
excited states, it would be very interesting to instead augment ESMF theory
by incorporating components from density functional theory while keeping the
formulation entirely time-independent.

%Besides the ability to describe weak correlations, the 
%state-specific orbital relaxation in ESMF also offers great advantages over 
%the widely used time-dependent density functional theory (TDDFT)\cite{Runge1984,casida2012,Ullrich2013,Burke2004}, in which the lacking of 
%excited state orbital relaxation could produce substantial amount of errors. 

%a DFT extension to ESMF method so that the missing dynamic correlations in ESMF excited 
%states could be recovered. Before we pursue further on this route, let us first 
%discuss the standard excited methods of KS-DFT.

%Density functional theory (DFT)\cite{Parr-Yang} based on the Kohn-Sham (KS) approach\cite{ks-dft} is the most widely used 
%electronic structure method in chemistry, physics, and materials science. Due to its
%favorable scaling with system size and reasonable accuracy in a variety of different
%circumstances, DFT is often regarded as one of the most powerful tools for studying 
%the electronic and dynamic properties of materials and medium to large molecules.
%Although originally formulated for ground states, the development of time-dependent density 
%functional theory (TDDFT)\cite{Runge1984, casida2012,Ullrich2013,Burke2004} extends DFT's capability to model excited states based on 

While the LR formulation of TDDFT is formally exact, approximations are needed
to make the approach computationally tractable.  
%TDDFT's lack of excited state orbital relaxation effects arises from the
%\textit{adiabatic approximation} (AA)\cite{Petersilka1995, Bauernschmitt1996}. 
The central quantity in TDDFT is the xc-kernel $f_{xc}(\textbf{r},t;\textbf{r}^{\prime},t^{\prime})$, 
defined as the functional derivative of the xc-potential,\cite{Petersilka1995}
\begin{equation}
    \label{eqn:xc_kernel}
    f_{xc}(\textbf{r},t;\textbf{r}^{\prime}t^{\prime})=\frac{\delta \upsilon_{xc}[n](\textbf{r},t)}{\delta n(\textbf{r}^{\prime},t^{\prime})}
\end{equation}
in which the $\upsilon_{xc}(\textbf{r},t)$ is the time-dependent analogy of the ground state xc-potential and $n(\textbf{r},t)$ is the electron density. 
The AA replaces the time-dependent xc-potential with the ground state 
xc-potential, \cite{Petersilka1995}
\begin{equation}
    \label{eqn:aa_vxc}
    \upsilon_{xc}^{adia}[n](\textbf{r},t)=\upsilon_{xc}^{GS}[n(t)](\textbf{r})
\end{equation}
at which point the xc-kernel becomes
\begin{equation}
    \label{eqn:aa_fxc}
    f_{xc}^{adia}(\textbf{r},t;\textbf{r}^{\prime}t^{\prime})=\frac{\delta \upsilon_{xc}^{GS}[n(t)](\textbf{r})}{\delta n(\textbf{r}^{\prime},t^{\prime})}=\delta(t-t^\prime)\frac{\delta^2E_{xc}[n]}{\delta n(\textbf{r})\delta n(\textbf{r}^\prime)}.
\end{equation}
Most notably, this approximation leads the xc-kernel to be local both in 
time and space if the ground state xc functional is local as in LDA,
or local in time but nonlocal in space in the case of hybrid functionals.

While the AA is enormously convenient in that it makes TDDFT efficient
and allows it to use existing ground state functionals,
it does create important limitations when modeling
%This makes TDDFT the most widely used 
%method to study the optical properties of medium to large size molecular systems. 
%For 
%TDDFT with AA has been used widely to compute optical properties of 
%The AA assumes that 
%the density changes slowly in the presence of an external, time-dependent field\cite{tddft2012}, which is the 
%driven force of electronic excitations. Such an assumption implies that the charge density 
%of the ground state and the excited states should be similar for the approximation to hold. 
%Consequently, when there are large amount of charge re-distributions between the ground 
charge transfer (CT), Rydberg, and double excitations.
%AA is no longer a physically sound approximation. 
For example, TDDFT 
often drastically underestimates excitation energies for long-range CT
states \cite{Dreuw2003,Nitta2012,Janus2013} and 
Rydberg states, \cite{Yang2013,Casida2000,VanMeer2014} and it is completely 
incapable of describing doubly excited states.\cite{Dreuw2005,Elliott2011} 
Besides the underestimation of excitation energies, it is also well known that for 
long-range CT excited states, standard pure and hybrid functionals also fail to exhibit the correct $1/R$ 
dependence along the charge separation coordination.\cite{Dreuw2003,Gritsenko2004} 
Given the technological and biological importance of CT, the limitations of
practical TDDFT in this area are especially frustrating.

To be more precise, these difficulties stem from two approximations: 
first, the usage of approximate xc functionals, and second, the AA. 
The former is responsible for the problems in Rydberg excited states 
and the missing $1/R$ behavior in long-range CT. 
For Rydberg states, the problem lies primarily in the fact that 
the ground state xc-potential of local and semi-local functionals decays 
exponentially with $r$, 
much faster than the $1/r$ decay of the exact xc-potential.
%Therefore they can not produce Rydberg series.
In order to deal with this problem, the asymptotic correction
approach \cite{Casida2000} has been developed and results in dramatically improved 
Rydberg energetics.
For CT excited states, range-separated hybrid functionals (RSH)
\cite{Chai2008, Chai2008a,Mardirossian2014,Mardirossian2016,Peverati2011,Lin2012}
successfully yield the correct $1/R$ behavior of long range CT excited states. 
%These functionals separate the Coloumb potential into a short-range and
%a long-range part and then gradually switches to a full Hartree-Fock (HF)
%exchange treatment in the long range limit.
This approach eliminates the CT self-interaction error in which derivatives
of the approximated exchange term fail to deliver the long range Coulomb term
that should be present in the linear response equations. \cite{Dreuw2005}
Even with the $1/R$ issue repaired, though, long range CT still poses challenges.
This is mainly due to fact that the excitation energy of long-range 
CT states should be determined by the ionization potential (IP) of the donor and 
the electron affinity (EA) of the acceptor.
While KS-DFT's highest occupied molecular orbital (HOMO) energy corresponds to the IP,
the lowest unoccupied molecular orbital (LUMO) energy does not, and is not supposed to,
correspond to the EA, even when the xc functional is exact. The result is that the 
difference between the DFT LUMO and HOMO energies severely underestimates the 
excitation energy, which leads to a situation in which the kernel contribution
to the TDDFT energy is asked to make up the difference.
However, the kernel contribution for most commonly used functionals is
typically much too small to make an appreciable difference on this scale,
and so CT energies get underestimated, sometimes quite severely.
This difficulty, which we will refer to as the EA/IP imbalance,
has been extensively studied \cite{Dreuw2003,Dreuw2005,Maitra2016,Maitra2017}
and can be seen clearly in the examples we investigate below.
%While we will discuss this issue further when analyzing our results, we will
%for now point out that substantial progress in this area has been made
%via the optimal tuning approach, \cite{baer2009otct,baer2012otfs}
%even if it is not always obvious how to select a single tuning parameter
%in cases where multiple different types of excitations are to be treated
%simultaneously.
%This issue is a reminder that delineations between Coulomb and exchange
%are an intuitive but ultimately artificial separation of the terms resulting
%from expectation values of the Coulomb operator.
%These delineations are often useful, but one should be careful not to read too
%much in to them, as TDDFT's issues with CT reminds us.

%that stems from missing nonlocal exchange and causes incorrect asymptotic
%behavior is repaired and the $1/R$ behavior is recovered. 
%(AC) approach\cite{Casida2000}%In terms of Rydberg excited 
%states, the accuracy of TDDFT can be dramatically improved by applying the asymptotic correction
%(AC) approach\cite{Casida2000}. 

Unlike the issues discussed above, TDDFT's failure to describe doubly
excited states can be laid squarely at the feet of the AA,
which converts the memory-dependent time-non-local
xc-kernel $f_{xc}(\textbf{r},t;\textbf{r}^{\prime}t^{\prime})$ into
a time-local affair with no memory.
One consequence of this simplification is that, when expressed in Fourier space,
the AA xc-kernel is frequency independent.
Given that it has been  shown that the exact xc-kernel carries a strong
frequency dependence near a double  excitation, \cite{Maitra2004,Maitra2016}
adiabatic xc-kernels are thus not appropriate or accurate for doubly excited states. 
In practice, the failure of the AA in describing doubly excited states also creates
difficulties for other excitations, especially in the context of CT.
%The incapability of describing doubly excited states of AA also has an impact on CT excited
%states.
As pointed out by Ziegler and coworkers, \cite{Ziegler2015,Ziegler2009}
another consequence of the AA is that it 
fails to account for relaxations in the occupied orbitals that are not
involved in the excitation.
The orbital shapes for the particle and hole are relaxed by TDDFT,
but the other orbital shapes are not, at least not when the AA is being used.
An intuitive way to see this in light of the double excitation limitation is to
consider that, after the single excitation itself, the leading order term in the
Taylor expansion of a fully orbital-relaxed singly excited state
is a linear combination of doubly excited determinants.
Since CT excited states undergo substantial charge deformations and
changes in dipole when compared to the ground state, allowing all of the orbitals
to relax during the excitation is crucial.\cite{Ziegler2008} 
Without full relaxation, errors in CT excitation energies
often reach multiple eVs, \cite{Dreuw2003} even when RSH functionals are employed.
In sum, it would be highly desirable to have an excited state methodology that
benefits from DFT's highly efficient incorporation of correlation effects but that
is free from the difficulties created by the AA and EA/IP imbalances.

In this paper, we present a density functional extension of the ESMF method (DFE-ESMF). 
Instead of relying on the linear response 
formalism and AA of TDDFT, we directly modify the energy expression of 
ESMF theory by borrowing key ingredients from KS-DFT.
While this does not lead to a formal density functional theory
as it lacks some of the key properties of ground state DFT,
the idea is to exploit density functionals' proven ability to
add weak correlation effects to an uncorrelated reference wave function.
As such correlations tend to be local, and a local region of a molecule
should not be capable of knowing whether it is formally part of an
excited state or a ground state, the hope is that the same ingredients
that allow KS-DFT to capture weak correlation effects will
remain effective in the excited state context.
As in the original formulation of ESMF, the energy expression
(including the newly incorporated DFT ingredients)
is combined with an excited state variational principle to achieve
excited-state-specific optimization of the orbitals. 
%Excited state variational principles have recently been successfully
%incorporated into both quantum Monte Carlo \cite{Zhao2016}
%and excited state generalizations of Hartree Fock theory.
%\cite{vanVoorhis2017sigmaSCF,Shea2018}
As we discuss below, this approach seeks to bypass both the
orbital relaxation and EA/IP imbalance issues that show up in
the practical application of TDDFT.
%This approach provides both a formally exact excited state DFT theory
%as well as a path to practical and accurate approximations
%via the generalization of KS's fictitious non-interacting
%ground state wave function into a fictitious minimally interacting
%excited state wave function.
%The result is a wave-function-based kinetic energy functional
%and a folding of all the non-classical contributions to the
%Coulomb energy (which are now a bit more complicated) into
%an approximate xc functional.
%Like the ground state DFT, the exact xc functional is unknown and approximate
%forms need to be used.
%Crucially, the variational nature of the approach
%allows the orbitals to relax in a state specific manner,
%and the theory simplifies to traditional KS theory when the ground
%state is targeted.
%As discussed, such a feature is crucial for CT excited states.
In a variety of exploratory calculations, we find that, when paired
with an xc functional with a high degree of exact exchange
(necessary to help alleviate a self-interaction bias stemming from
excited states' more open-shell character),
this DFE-ESMF approach provides an accuracy comparable to TDDFT for
simple single-configuration-state-funciton (single-CSF) valence
excitations while far outperforming
it in CT states, even when comparing against a RSH functional.
The performance for multi-CSF single excitations
is more mixed, which appears to be caused by double-counting issues
as we discuss in some detail below.

This paper is organized as follows.
We begin with a brief review of ground state KS-DFT so as to make
clear its parallels with our excited state formalism.
We then develop the working equations for the DFE-ESMF method
in the context of both single-configurational and
multi-configurational wave functions.
We then briefly review the ground state xc functionals that
we employ and discuss concerns about possible double counting problems.
At the end of the theory section, we compare DFE-ESMF with other 
excited state and multi-reference DFT methods
and also with constrained DFT.
Results and discussions are then presented for a variety of different
valence, CT, and Rydberg excitations.
We conclude our discussion by pointing out the merits and 
drawbacks of the current method, along with possible directions
for future development. 

\section{Theory}
\label{sec:theory}

\subsection{Ground State DFT}
\label{sec:ksdft}

In ground state DFT, the Levy constrained search formulation provides
a formally exact energy functional, \cite{Parr-Yang}
\begin{equation}
    \label{eqn:levy_gs}
    E[n]=\underset{\Psi\rightarrow n}{\textrm{min}}\left<\Psi|\hat{T}+\hat{V}_{ee}|\Psi\right>+V_{ext}[n]
\end{equation}
in which $\hat{T}$ and $\hat{V}_{ee}$ are the kinetic and electron-electron
repulsion operators, and $V_{ext}[n]$ is the external potential.
In practice, KS-DFT re-writes this functional as
\cite{Parr-Yang,ks-dft}
\begin{equation}
\label{eqn:tot_eng}
E[n]=T_s[n]+V_{ext}[n]+J[n]+E_{xc}[n],
\end{equation}
in which $T_s[n]$ is the kinetic energy of a fictitious Slater determinant
that shares the same density as the actual interacting system,
$J[n]$ is the Hartree part of the electron-electron repulsion energy,
and $E_{xc}[n]$ is the exchange-correlation functional. 
Considering the common case of a closed-shell, spin-restricted KS determinant
for $N$ electrons, one can re-express the energy in terms of the orbitals
$\phi_i(\textbf{r})$ for $i\in [1,2,\ldots,N/2]$.
The external potential and Hartree pieces,
\begin{align}
    \label{eqn:ext_pot}
    V_{ext}[n]&=\int{\upsilon_{ext}(\textbf{r})n(\textbf{r})d\textbf{r}} \\
    \label{eqn:jofrho}
    J[n]&=\frac{1}{2}\int{\frac{n(\textbf{r})n(\textbf{r}^{\prime})}{|\textbf{r}-\textbf{r}^{\prime}|}d\textbf{r}d\textbf{r}^{\prime}} 
\end{align}
are dependent only on the density, which is in turn now determined by
the orbitals\cite{ks-dft},
\begin{equation}
    \label{eqn:gs_den}
    n(\textbf{r})=2\sum_{i}^{N/2}|\phi_i(\textbf{r})|^2.
\end{equation} 
Note that we follow the convention that $i,j,k$ refer to occupied orbitals, $a,b,c$ to virtual orbitals, and $p,q,r,s$ to all orbitals.
In the original KS formulation\cite{ks-dft}, the xc functional depends only on
density. However, due to the development of generalized KS schemes\cite{Becke1993gks,Levy1995} and hybrid functionals, it becomes more 
appropriate to write the xc function as a direct function of the orbitals,
\begin{align}
\label{eqn:exc}
E_{xc}[n] \rightarrow E_{xc}\big(\phi_1, \phi_2, \ldots, \phi_{N/2}\big).
\end{align}
Of course, the KS kinetic energy is also an
orbital functional,
\begin{equation}
\label{eqn:k_eng}
T_s[n] =
% \left<\Psi_s|\hat{T}|\Psi_s\right> =
-\sum_{i}^{N/2}{\int{\phi_i(\textbf{r})\nabla^{2}\phi_i(\textbf{r})d\textbf{r}}}.
\end{equation}
With this orbital-based formulation, one then minimizes
Eq.\ (\ref{eqn:tot_eng}) under the constraint that the
orbitals remain orthonormal in order to
arrive at the KS orbital eigenvector equation,
\begin{equation}
\label{eqn:ks_oe}
\hat{F}_{{}_{KS}} \hspace{0.5mm} \phi_i =
\epsilon_i \hspace{0.5mm} \phi_i,
\end{equation}
in which $\hat{F}_{{}_{KS}}$ is the KS Fock operator.

Crucially, we note that the same energies, orbitals, and densities
are arrived at if one performs the minimization
\begin{align}
\label{eqn:ks_eng}
E_{{}_{KS}} =
\underset{\bm{X}}{\textrm{min}}
\Big\{
T_s + V_{ext} + J + E_{xc}
\Big\}
\end{align}
in terms of the elements of the anti-Hermitian matrix
$\bm{X}$ that transforms some initial orthonormal set of orbitals
(such as those that diagonalize the one-electron parts of $\hat{H}$)
into the final KS orbitals.
\begin{align}
\label{eqn:mo}
\phi_p(\textbf{r})
\hspace{0.3mm} = \hspace{0.7mm}
%\sum_q \mathrm{exp}(\bm{X})_{pq}
\sum_q \left[e^{\bm{X}}\right]_{pq}
\hspace{0.7mm} \phi^{(0)}_q(\textbf{r})
\end{align}

Given ESMF's similarities to HF, it is worthwhile to write the HF energy in this same form.
%In order to make the discussion below clear, we also write down the energy minimization
%of HF method,
\begin{align}
    \label{eqn:hf_eng}
    E_{{}_{HF}} =
\underset{\bm{X}}{\textrm{min}}
\Big\{
T_s + V_{ext} + J + E_{x}
\Big\}
\end{align}
Here $E_x$ is the HF exchange energy
\begin{equation}
    \label{eqn:ex_gs}
    E_x=-\sum_{ij}{(ij|ji)}
\end{equation}
which we have expressed in terms of the two-electron integrals in the relaxed orbital basis.
\begin{equation}
    \label{eqn:2ei}
    (pq|rs) = \int{ \int{
     \frac{ \phi_p(\textbf{r}) \phi_q(\textbf{r})
            \phi_r(\textbf{r}^\prime) \phi_s(\textbf{r}^\prime)
          }{|\textbf{r}-\textbf{r}^\prime|}
     d\textbf{r} d\textbf{r}^\prime
    } }
\end{equation}
By comparing Equation \ref{eqn:ks_eng} and \ref{eqn:hf_eng}, KS-DFT can be seen as the pairing of
a minimally-correlated ansatz (the Slater determinant)
and a variational principle (the total energy)
in which the energy expression within the latter
has been augmented by modifying the exchange term.
To formulate DFE-ESMF, we will follow a similar route,
but with the ESMF wave function as the minimally-correlated ansatz and using an excited state variational principle instead of simple energy minimization.

\subsection{DFE-ESMF: Single-CSF Formalism}
\label{sec:single_csf}

In ESMF, excited states are targeted by applying the following Lagrangian 
form of an excited state variational principle,\cite{Shea2018}
%Take the variational principle first.
%Drawing on the strong parallels between HF and DFT, we will
%import the Lagrangian form of the excited state variational principle
%used in the ESMF theory, \cite{Shea2018}
%which is a generalization of HF theory for excited states.
\begin{align}
\label{eqn:esmf_exact_L}
L =  \left<\Psi|(\omega-\hat{H})^2|\Psi\right>
     - \bm{\mu} \cdot \frac{\partial E}{\partial \bm{\nu}}
\end{align}
Here $\omega$ is an energy used to select which excited state is being
targeted, $\bm{\nu}$ is the vector of variational parameters within
the ansatz, and $\bm{\mu}$ is a vector of Lagrange multipliers
by which we constrain the the minimization of $L$ so that it must
converge to an energy stationary point.
In essence, the first term in $L$ is a rigorous excited state variational
principle with the energy eigenstate closest to $\omega$ as its global
minimum, but because approximate ansatzes will prevent us from reaching
this minimum, we add the energy stationarity constraint to ensure that at
least this important property of exact excited states is maintained.
In other words, the idea is for the first term to drive the optimization
to the energy stationary point that best corresponds to the desired excited state.
In preliminary work on ESMF, \cite{Shea2018} it has been found that
computationally tractable approximations to this Lagrangian
\begin{align}
\label{eqn:esmf_approx_L}
\widetilde{L} =  (\omega-E)^2
     - \bm{\mu} \cdot \frac{\partial E}{\partial \bm{\nu}}
\end{align}
are in practice effective at achieving the same goal, and so for
expediency's sake we will adopt $\widetilde{L}$ as our working variational
principle for DFE-ESMF.

Before augmenting the ESMF energy expression with density functional components, we should consider the choice of the wave function ansatz used in ESMF.
To start, consider a singly excited configuration state function (CSF),
which is perhaps the simplest spin-pure excited state ansatz.
\begin{equation}
 \label{eqn:csf}
 \left|\Psi_i^a\right>
 %(i\rightarrow a)(X)\right>
 %=\frac{1}{\sqrt{2}}\Big(\left|\Psi_i^a(X)\right>+\left|\Psi_{\bar{i}}^{\bar{a}}(X)\right>\Big)
 = \frac{1}{\sqrt{2}}\Big(
    a^+_a a_i \left|\Psi_0\right> \pm a^+_{\bar{a}} a_{\bar{i}} \left|\Psi_0\right>
   \Big)
\end{equation}
This is a superposition between alpha ($i\rightarrow a$) and
beta ($\bar{i}\rightarrow\bar{a}$) excitations from the
closed shell Slater determinant $\Psi_0$ in which the excitations
both occur within the same pair of spatial orbitals $\{i,a\}$.
The sign determines whether the excitation is a singlet or triplet,
and although we will develop the mathematics for the singlet case below,
the triplet is equally straightforward.
%has e ground state has been excited by promoting an electron from the $i$th occupied orbital to the $a$th virtual orbital. 
%$\left|\Psi_i^a\right>$ represents an excited determinant in which an alpha electron is promoted,
%%the $i$th occupied orbital to the $a$th virtual orbital,
%and $\left|\Psi_{\bar{i}}^{\bar{a}}\right>$ is the 
%corresponding beta-excited determinant.
%For simplicity of presentation, we will develop the theory for the singlet case,
%but note that the derivation is essentially the same for triplet states. 
As in the ground state presentation above, the (spin-restricted) molecular orbitals will
be defined via an anti-Hermitian matrix $\bm{X}$ as in Eq.\ (\ref{eqn:mo}),
but with the corresponding ground state KS-DFT orbitals now acting as the
initial orbitals $\phi^{(0)}$ and $\bm{X}$ encoding excited-state-specific
relaxations.
This single-CSF ansatz leads to the electron density
%
%In Equation \ref{eqn:csf}, the orbitals indexed by $i$ and $a$ are not necessarily the KS orbitals
%since they are allowed to relax from the starting orbitals (usually taken to be the KS orbitals)
%through the anti-hermitian matrix X:
%
%\begin{equation}
%    \label{eqn:x}
%    \phi_p(\textbf{r})(X)=\sum_{q}{(e^X)_{pq}\phi^{KS}_q(\textbf{r})}
%\end{equation}
%In the current study we use spin-restricted real orbitals. 
%The density for this singly-excited fictitious wave function is:
\begin{equation}
  \label{eqn:exden}
  n_{ia}(\textbf{r}) =
    \left|\phi_a(\textbf{r})\right|^2
     - \left|\phi_i(\textbf{r})\right|^2
     + 2\sum_k^{N/2} \left|\phi_k(\textbf{r})\right|^2
\end{equation}
and a one-body reduced density matrix (1RDM)
$\bm{P}$ that is diagonal in
the basis of the relaxed molecular orbitals.
\begin{align}
  \notag
  P_{kj}&=2\delta_{kj}-\delta_{ki}\delta_{ji} \\
  \label{eqn:1rdm} 
  P_{bc}&=\delta_{ba}\delta_{ca} \\
  \notag
  P_{jb}&=P_{bj}=0
\end{align}
%As in KS-DFT, we set the kinetic energy to that of the ESMF
%minimally correlated wave function.
%We express this in the atomic orbital (AO) basis by
%rotating the 1RDM in to that basis,
%\begin{equation}
%    \label{eqn:1rdm_atom}
%    \bm{P}_{{}_{AO}} = \bm{C}_{{}_{KS}} e^{\bm{X}} \bm{P} e^{-\bm{X}} %\bm{C}^+_{{}_{KS}}
%\end{equation}
The ESMF kinetic energy can be computed as
%where we have used the KS orbital coefficient matrix $\bm{C}_{{}_{KS}}$.
%The kinetic energy is then
\begin{equation}
    \label{eqn:ex_kine}
        T_{ia} = \left<\Psi_i^a|\hat{T}|\Psi_i^a\right>
        = \mathrm{Tr}\left[\bm{P}_{{}_{AO}} \bm{T}_{{}_{AO}} \right]
\end{equation}
where $\bm{P}_{AO}$ is the 1RDM rotated into the atomic orbital basis, and $\bm{T}_{{}_{AO}}$ are the kinetic energy integrals in that basis.
%\begin{align}
%  \begin{split}
%       &\left[\bm{T}_{AO}\right]_{pq}
%    = -\frac{1}{2}\int{\chi_p(\textbf{r})\nabla^2\chi_q(\textbf{r})d\textbf{r}}, \\
%    &\bm{P}_{{}_{AO}} = \bm{C}_{{}_{KS}} e^{\bm{X}} \bm{P} e^{-\bm{X}}\bm{C}^+_{{}_{KS}}
%  \end{split}
%\end{align}
%and $\chi_p$ are the AO basis functions. 
%By similarly converting the expression for the density
%in to the AO basis,
%Now we can write the density in terms the 1RDMs in atomic orbital basis:
%\begin{equation}
%    \label{eqn:den_1rdm}
%    n_{ia}(\textbf{r}) =
%      \sum_{pq} \chi_p(\textbf{r})
%                \left[\bm{P}_{{}_{AO}}\right]_{pq}
%                \chi_q(\textbf{r}),
%\end{equation}
%in which $\psi_p(\textbf{r})$ is the $p$th atomic orbital evaluated at position $\textbf{r}$. 
%With the aid of 1RDMs, we can also compute the kinetic energy of the
%wave function in Equation \ref{eqn:csf}.  
Likewise, the external potential contribution may be evaluated as
%, which is the electron-nuclei attraction energy in the current study, can
%also be expressed in terms of the 1RDMs:
\begin{align}
\label{eqn:en_att}
V^{ext}_{ia} &= \mathrm{Tr}\left[ \bm{P}_{{}_{AO}} \bm{h}_{{}_{AO}} \right]
%  \left[\bm{h}_{AO}\right]_{pq}
%  &= \int{ \chi_p(\textbf{r})\upsilon_{ext}(\textbf{r})\chi_q(\textbf{r}) d\textbf{r} }.
%    \int{d\textbf{r}\upsilon_{ext}(\textbf{r})n_{ia}(\textbf{r})} \\
%    &=\sum_{pq}{\tilde{P}_{pq}\int{d\textbf{r}\psi_p(\textbf{r})\upsilon_{ext}(\textbf{r})\psi_q(\textbf{r})}} =\sum_{pq}{\tilde{P}_{pq}h_{pq}}
\end{align}
%in which we define $h_{pq}=\int{d\textbf{r}\psi_p(\textbf{r})\upsilon_{ext}(\textbf{r})\psi_q(\textbf{r})}$. 
where $\bm{h}_{AO}$ are the corresponding one-electron integrals.

Turning our attention now to the electron-electron repulsion energy,
we start with the Hartree term, which for this singlet CSF's density is \\
\begin{align}
    \notag
    J[n_{ia}]
    %\frac{1}{2}\int{d\textbf{r}d\textbf{r}^\prime\frac{n_{ia}(\textbf{r})n_{ia}(\textbf{r}^\prime)}{|\textbf{r}-\textbf{r}^\prime|}} \\
    \hspace{0.5mm} = \hspace{1.0mm} & \frac{1}{2}(ii|ii)
    \hspace{0.5mm} + \hspace{0.5mm}   \frac{1}{2}(aa|aa)
    \hspace{0.5mm} - \hspace{0.5mm}   (aa|ii) \\
    & \hspace{1.0mm} + \hspace{0.5mm} 2\sum_{kj}(kk|jj)
      \hspace{0.5mm} + \hspace{0.5mm} 2\sum_j{ \Big( (aa|jj) - (ii|jj) \Big) },
    \label{eqn:classic_J}
\end{align}
%which we have expressed in terms of the two-electron integrals in the relaxed orbital basis.
%\begin{equation}
%    \label{eqn:2ei}
%    (pq|rs) = \int{ \int{
%     \frac{ \phi_p(\textbf{r}) \phi_q(\textbf{r})
%            \phi_r(\textbf{r}^\prime) \phi_s(\textbf{r}^\prime)
%          }{|\textbf{r}-\textbf{r}^\prime|}
%     d\textbf{r} d\textbf{r}^\prime
%    } }
%\end{equation}
For hybrid functionals, we will need a definition for the 
wave-function-based exchange energy (i.e.\ an excited state analogue of HF exchange), which we choose to
arrive at by making the usual index exchanges in the two-electron integrals
of the corresponding Hartree term.
%We can also compute an exchange energy by making the usual index exchange in the $J_{ia}[n]$ formula,
\begin{align}
\notag
    E^{x\hspace{0.2mm}(\mathrm{wfn})}_{ia}
    \hspace{0.5mm} = \hspace{1.0mm} & -\frac{1}{2}(ii|ii)
    \hspace{0.5mm} - \hspace{0.5mm}   \frac{1}{2}(aa|aa)
    \hspace{0.5mm} + \hspace{0.5mm}   (ai|ia) \\
    & \hspace{1.0mm} - \hspace{0.5mm} \sum_{kj}(kj|jk)
      \hspace{0.5mm} - \hspace{0.5mm} \sum_j{ \Big( (aj|ja) - (ij|ji) \Big) }
% = -\sum_{ij}(ij|ji)+\sum_j{[-(ja|aj)+(ji|ij)]} \\
%&+(ia|ai)-\frac{1}{2}(ii|ii)-\frac{1}{2}(aa|aa)
\label{eqn:exx}
\end{align} 
%
%The exchange-correlation energy is computed in the same way as in the ground state DFT,
%\begin{equation}
%    \label{eqn:ex_exc}
%    E_{xc}=\int{d\textbf{r}\upsilon_{xc}(\textbf{r})n(\textbf{r})}
%\end{equation}
%in which the specific form of $\upsilon_{xc}(\textbf{r})$ depends on the approximation form (LDA, GGA\cite{Perdew1992, Perdew1996a}...)
%being used. 
%
Now, for the closed shell Slater determinant used in KS-DFT,
the Hartree term (Equation \ref{eqn:jofrho}) and the wave-function exchange term (Equation \ref{eqn:ex_gs}) sum to the electron-electron repulsion
energy of the Slater determinant.
However, things are not so simple in the excited state, and even for this
single-CSF singlet wave function, the full wave-function-based electron-electron
repulsion energy contains one additional term:
%In the ground state DFT, the classical Coulomb energy $J_{ia}[n]$ and the ``exact'' exchange
%energy defined by these index exchanges sum exactly to the full quantum mechanical 
%electron-electron repulsion energy of the fictitious KS determinant. 
%However, adding the Coulomb energy in Equation
%\ref{eqn:classic_J} with the exchange energy in Equation \ref{eqn:exx} does not yield 
%the full quantum mechanical electron-electron repulsion energy $E^{ee}_{ia}$ energy of the wave function in Equation \ref{eqn:csf}, which is instead:
\begin{align}
    E^{ee}_{ia}
    \hspace{0.3mm} \equiv \hspace{0.3mm} \left<\Psi^a_i\right| \hat{V}_{ee} \left|\Psi^a_i\right>
    \hspace{0.5mm} = \hspace{0.5mm} J[n_{ia}]
    \hspace{0.5mm} + \hspace{0.5mm} E^{x\hspace{0.2mm}(\mathrm{wfn})}_{ia}
    \hspace{0.1mm} + \hspace{0.5mm} (ai|ia).
    \label{eqn:csf_2b}
\end{align}
For the triplet CSF, we have a similar situation, but the additional term
takes on the opposite sign:
\begin{align}
    E^{ee \hspace{0.2mm}(\mathrm{triplet})}_{ia}
%    \hspace{0.3mm} \equiv \hspace{0.3mm} \left<\Psi^a_i\right| \hat{V}_{ee} \left|\Psi^a_i\right>
    \hspace{0.5mm} = \hspace{0.5mm} J[n_{ia}]
    \hspace{0.5mm} + \hspace{0.5mm} E^{x\hspace{0.2mm}(\mathrm{wfn})}_{ia}
    \hspace{0.1mm} - \hspace{0.5mm} (ai|ia).
    \label{eqn:csf_triplet_sb}
\end{align}
This extra term, which we will denote as the wave function correlation energy (WFCE)
\begin{align}
    E^{\mathrm{WFCE}}_{ia} = \pm (ai|ia),
\end{align}
determines the singlet-triplet splitting and arises
from the fact that $\hat{V}_{ee}$ connects the two different terms in our CSF.

The energy of the wave function in Equation \ref{eqn:csf} is then the sum of the kinetic energy,
the external potential, Hartree and exchange energy, and WFCE:
\begin{equation}
    \label{eqn:wfn_eng_csf}
    E^{\mathrm{wfn}}_{ia}=T_{ia}+V^{ext}_{ia}+J[n_{ia}]+E^{x(\mathrm{wfn})}_{ia}+E^{\mathrm{WFCE}}_{ia}.
\end{equation}
In the same way that one can arrive at KS-DFT starting from HF by replacing the exchange term with an exchange correlation functional (which converts Equation \ref{eqn:hf_eng} into Equation \ref{eqn:ks_eng}), we now replace the exchange term in our excited state wave function energy in order to arrive at the energy expression
for single-CSF DFE-ESMF.
\begin{equation}
    \label{eqn:exDFT_eng}
    E_{ia} =
       T_{ia}
     + V^{ext}_{ia}
     + J[n_{ia}]
     + E^{xc}_{ia}
     + E^{\mathrm{WFCE}}_{ia}
\end{equation}
Note especially that the WFCE term is retained.
As this term originates from a strong correlation effect
(the two electrons involved in the excitation are taking care
to never be in the same orbital at the same time),
we assume that it will not create significant double counting issues
when used in conjunction with standard formulations of ground state
exchange-correlation functionals, as these are geared towards weak
correlation and are not designed to capture open-shell spin recoupling correlations.
In ground state KS-DFT, $E^{\mathrm{WFCE}}=0$ and
HF is recovered by using a functional with no correlation and 
100\% HF exchange. The analogous property is maintained by DFE-ESMF: 
when using a functional consisting soley of 100\% wave function exchange
as defined in Equation \ref{eqn:exx}, 
the DFE-ESMF energy reverts back to the ESMF expression for a single CSF's
energy.
%Thus, in combining Eqs.\ (\ref{eqn:esmf_approx_L}) and (\ref{eqn:exDFT_eng}),
%we followed the same general approach that leads to 
%KS-DFT in the ground state case.
%Specifically, we combined a minimally correlated ansatz
%with a variational principle in which the energy was
%replaced with that of a density functional.

As in ESMF theory, the minimization of Eq.\ (\ref{eqn:esmf_approx_L}) requires
the evaluation of certain sums over the second derivatives of our
energy expression.
Although the density functional energy expression of
Eq.\ (\ref{eqn:exDFT_eng}) differs from that of ESMF theory, 
we can exploit the same automatic differentiation (AD) approach
in order to perform the optimization at a cost whose scaling with
system size is the same as a ground state KS Fock build.
For an explanation of how this is achieved, we refer the reader
to the original ESMF paper. \cite{Shea2018}
As in that case, we have formulated our pilot code
using the convenient AD capabilities of the
TensorFlow framework \cite{tensorflow2015-whitepaper}
and have carried out the minimization via
a quasi-Newton approach. \cite{Nocedal1980}
In addition to what is necessary for ESMF, this requires AD through
the grid integration involved in density functional
components such as the LDA exchange and correlation terms,
which we have now achieved with the correct scaling.

\subsection{DFE-ESMF: Multiple-CSF Formalism}
\label{sec:multi_csf}

In cases where a state contains major contributions from multiple
different single excitations, we may generalize the approach
into a multi-CSF form with a wave function similar to
%we need to generalize the wave function in Equation \ref{eqn:csf} to a 
configuration interaction singles (CIS),\cite{Hirata1999}
\begin{equation}
    \label{eqn:cis}
    \left|\Psi_{\textrm{MCSF}}\right>=\sum_{ia}{c_{ia}\left|\Psi_i^a\right>},
\end{equation}
in which we still relax the orbitals as above.
In this case, the density becomes
%{\color{red} (check this Chris)}
\begin{align}
  \notag
  n_{{}_{\textrm{MCSF}}}(\textbf{r}) =
  4\sum_{ia}{ |c_{ia}|^2 \sum_{k}{ \left|\phi_k(\textbf{r})\right|^2 }} 
   &+ 2\sum_{iab}{c_{ia}c_{ib}\phi_a(\textbf{r})\phi_b(\textbf{r})} \\
   &- 2\sum_{ija}{c_{ia}c_{ja}\phi_i(\textbf{r})\phi_j(\textbf{r})}
\label{eqn:cis_den}
\end{align}
and the 1RDM in the relaxed MO basis is no longer diagonal.
\begin{equation}
    \label{eqn:cis_1rdm}
    \begin{split}
        &P_{ij}=\delta_{ij}-\sum_{a}{c_{ia}c_{ja}} \\
        &P_{ia}=P_{ai}=0 \\
        &P_{ab}=\sum_{i}{c_{ia}c_{ib}} \\
    \end{split}
\end{equation}
Nonetheless, we can still take the KS approach and evaluate both the kinetic
energy and external potential via the wave function's 1RDM
using Eqs.\ (\ref{eqn:ex_kine}) and (\ref{eqn:en_att}). 

Although the one-electron components are quite similar to the single-CSF
approach, the electron-electron repulsion energy is less straightforward.
In order to define the Hartree term, one possibility is to use the density 
from Eq.\ (\ref{eqn:cis_den}) in the standard $J[n]$ form of Eq.\ (\ref{eqn:jofrho}).
However,  doing so introduces unphysical virtual-virtual Coulomb repulsion terms 
in the form of $(aa|bb)$, similar to the ghost interactions encountered in
ensemble DFT.
In order to avoid these in the multi-CSF case, we generalize the Hartree
term as the weighted  statistical average of the Hartree terms
from each separate CSF as given in Eq.\ (\ref{eqn:classic_J}).
\begin{align}
  \label{eqn:cis_coul}
  J_{\textrm{MCSF}} \equiv& \sum_{ia}{|c_{ia}|^2 J[n_{ia}]}
%=& 2\sum_{ij}{(ii|jj)} + \sum_{ia}{C^2_{ia}\sum_j{[4(aa|jj)-4(ii|jj)]}} \\ &+\sum_{ia}{C^2_{ia}[-2(aa|ii)+(ii|ii)+(aa|aa)]} \\
\end{align}
If we now apply the index-exchange approach, we simply arrive
at an ``exact'' wave function exchange that is the weighted
average of the single-CSF pieces from Eq.\ (\ref{eqn:exx}).
\begin{equation}
\label{eqn:cis_exx}
E^{x\hspace{0.2mm}(\mathrm{wfn})}_{MCSF}
\equiv \sum_{ia}{ |c_{ia}|^2 E^{x\hspace{0.2mm}(\mathrm{wfn})}_{ia} }
%&=-\sum_{ij}{(ij|ji)}+\sum_{ia}{C^2_{ia}\sum_j{[-2(ja|aj)+2(ji|ij)]}} \\
%&+\sum_{ia}{C^2_{ia}[2(ia|ai)-(ii|ii)-(aa|aa)]} \\
\end{equation}
As before, the Hartree and exchange pieces do not add up to the full
wave function electron-electron repulsion energy,
\begin{align}
\left<\Psi\right| \hat{V}_{ee} \left|\Psi\right>
 = J_{\textrm{MCSF}} + E^{x\hspace{0.2mm}(\mathrm{wfn})}_{\mathrm{MCSF}}
   + E^{\mathrm{WFCE}}_{\textrm{MCSF}},
\end{align}
and the additional correlation effects are now more involved.
\begin{equation}
\label{eqn:cis_corr}
\begin{split}
    &E^{\mathrm{WFCE}}_{\textrm{MCSF}}
    =2\sum_{iajb}{c_{ia}c_{jb}[2(ai|jb)-(ab|ji)]} \\
    &+\sum_{abi}{c_{ia}c_{jb}[\sum_k{4(ab|kk)-2(ak|kb)}]} \\
    &-\sum_{ija}{c_{ia}c_{ja}[\sum_k{4(ij|kk)-2(jk|ki)}]} \\
    &+\sum_{ia}{|c_{ia}|^2\sum_k{[-4(aa|kk)+2(ak|ka)+4(ii|kk)-2(ik|ki)]}} \\
    &+\sum_{ia}{|c_{ia}|^2[2(aa|ii)-2(ai|ia)]} \\
\end{split}
\end{equation}
%Clearly, the FSCE is now a lot more complicated 
%compared to a single term in the single-CSF formalism. 

Using the same logic as before (although see Section
\ref{sec:dcp} regarding double counting concerns),
we define the multi-CSF density functional form for the energy in 
Eq.\ (\ref{eqn:esmf_approx_L}) to be 
\begin{equation}
    \label{eqn:exDFT_mcsf}
    E_{\mathrm{MCSF}} = T + V_{ext} + J_{\mathrm{MCSF}} + E_{xc}
    + E^{\mathrm{WFCE}}_{\textrm{MCSF}}
\end{equation}
in which the $T$ and $V_{ext}$ are as in Eqs.\ (\ref{eqn:ex_kine})
and (\ref{eqn:en_att}) but with the multi-CSF 1RDM,
and $E_{xc}$ is as in the ground state functional but with the density
taken from Eq.\ (\ref{eqn:cis_den}) and the wave function exchange component
set to $E^{x\hspace{0.2mm}(\mathrm{wfn})}_{\mathrm{MCSF}}$.
%, $J[n]$, and $E_{FSCE}$ are defined 
%in Equation \ref{eqn:ex_kine}, \ref{eqn:en_att}, \ref{eqn:cis_coul},
%and \ref{eqn:cis_corr} respectively. 

%Although this definition is not unique due to our somewhat arbitrary choice
%for the Hartree and exchange terms, it does offer three advantages.
%First, unphysical virtual-virtual repulsions are avoided.
%Second, as in both KS-DFT and our single-CSF formalism, setting the xc
%functional to 100\% exact exchange with no correlation
%recovers the wave function theory, which in this case is ESMF theory.
%\cite{Shea2018} 
%Third, as in KS-DFT, each term that results from expanding
%Eq.\ (\ref{eqn:cis_coul}) has a direct classical interpretation.

\subsection{Double Counting Problems}
\label{sec:dcp}

The DFE-ESMF energy, in both the single and multiple CSF formalisms, contains 
correlation terms that do not exist in the energy expression of ground state DFT.
However, one potential problem of adding these correlation terms into the 
energy formula as we have done is that, in principle, they could be accounted for
again in the xc functional, leading to a double counting problem.
In the single-CSF formalism, the WFCE term $(ai|ia)$ arises completely due to 
the fact that the wave function contains two determinants with equal weights. 
%Otherwise, the overall spin-symmetry would be broken and singlet-triplet gaps incorrect.
Such a strong correlation
effect is (typically) not built in to practical forms for $E_{xc}$ 
which instead aim to include weak correlation effects. \cite{Grafenstein2000}
Therefore, we do not expect to have significant double counting problems in the single-CSF case. 

However, if one employs the full CIS-style multi-CSF formalism, the wave function
definitely includes both some kinetic energy correlation effects and some
electron-electron interaction correlation effects.
In order to illustrate this, consider the case where the multi-CSF expansion is 
dominated by one CSF with an excitation between the $i$th and $a$th orbitals.
We can treat this dominant piece as the zeroth-order reference in a perturbative
expansion.
As some other singly-excited CSFs are coupled to this reference
by $\hat{T}$ and even more by $\hat{V}_{ee}$,
such couplings would be part of any 2nd-order perturbation
correction starting from this reference.
Thus, a simple Moller-Plesset-style argument suggests that many and perhaps most of
the contributions within $E^{\mathrm{WFCE}}_{\textrm{MCSF}}$ would be part of the
system's weak correlation physics and so at significant risk of double counting
within our multi-CSF formalism.
Indeed, in early testing, we found that excitation energies with the full
multi-CSF formalism were worse than those from the single-CSF formalism, which
we now understand was primarily a double counting issue.

In order to avoid this problem, one might try to separate contributions 
from the weak and strong correlations within the multi-CSF expansion.
Although there is no unique way to do this, we have for now taken the expedient
approach of including in our multi-CSF expansion only those CSFs
whose TDDFT coefficients are above a threshold.
While it may become clear once more data is available what the least-bad
threshold choice would be, we have for now set this threshold at a relatively large
value of 0.2 to help ensure that retained CSFs are playing a larger-than-perturbative
role in the excitation and are therefore more likely to contribute energetic 
correlation effects of the type that are not built in to common density functionals.
For simplicity, and in contrast to ESMF theory, we do not optimize these coefficients
in our minimization of $\widetilde{L}$ and instead hold them fixed at their TDDFT values. Admittedly, such a 0.2 threshold will become troublesome in cases 
where the excited states are composed of a large collection of excitations, such as 
plasmons. Therefore, developing alternative approaches to avoid the double counting 
problem will be highly desired in future developments of DFE-ESMF. 

\subsection{Discussion of DFE-ESMF}
It is important to note that the DFE-ESMF method in its current form 
is not an excited state generalization of the ground state KS-DFT. Based on 
the Hohenberg-Kohn theorem\cite{Parr-Yang}, which establishes a one-to-one 
mapping between external potential and density, the ground state energy depends
solely on density. However, it has been shown\cite{Gaudoin1004} that such
a one-to-one mapping between external potential and density does not exist for excited states. Therefore, the excited state density alone can not uniquely 
determine its energy. The simplest example would the singlet and triplet excited state
of a given configuration. These two states have the same density, but different 
energies. In previous developments that try to generalize the ground state 
KS-DFT to excited states, Levy and Nagy use bi-functionals\cite{Nagy1999} that
depends on both excited state and ground state density, and G\"orling uses totally 
symmetric part of the density\cite{Gorling1993} and a generalized adiabatic connection scheme\cite{Gorling1999, Gorling2000}, in order to enforce the correct symmetry
of excited states.
Thus, we suggest that it is more useful to view
DFE-ESMF as a practical extension to ESMF rather than as a formal density
functional theory.
That said, DFE-ESMF does share some similarities with the exact
generalized adiabatic connection (GAC) approach. \cite{Gorling2000}
Both methods use a symmetry-determined linear combination of Slater determinant
to compute kinetic energy, external potential, and exchange energy. 
%Furthermore, in the initial 
%implementation of GAC, total density, rather than the symmetric part of the 
%density, is used in the energy functional. This adds another layer of similarities 
%between these two methods. 
In addition, both methods try to enforce the correct excited state symmetry. 
In DFE-ESMF, the excited state symmetry is taken
care by the WFCE term, while GAC uses the symmetrized density.\cite{Gorling2000}

In the context of spin symmetry, it is important to note that the
WFCE term is essential for our optimization approach.
Without this term, singlet and triplet excited states formulated by the same 
excitation have the same density and energy. Consequently, our energy-based
excited state variational approach would not be able to
distinguish these two states and its results would be arbitrary.
As discussed before, the difference between singlet and triplet 
states is encoded in the WFCE term, and adding this term to the energy expression 
greatly helps the optimization procedure to pick the desired state. 

At present, DFE-ESMF uses functionals developed for ground state to treat 
excited states.
While there is no reason to think that this approach is
optimal, it is a very common procedure to treat excited states using
ground state functionals.
For example, the aforementioned 
GAC approach, the $\Delta$ self-consistent field ($\Delta$SCF) method, \cite{Ziegler1997}
constricted variational density functional theory, \cite{df-excited}
and spin-restricted ensemble-referenced Kohn-Sham method (REKS) \cite{Filatov2015}
all use ground state functionals to describe excited states. 
While it may in future be worthwhile to develop functionals specifically for use
with DFE-ESMF, we do not explore this direction here.

Although we do use ground state functionals, it is important to distinguish
the present approach from the use of such functionals in TDDFT via the AA.
First, the AA is a statement about the time dependence of the exchange correlation
kernel, which has no direct analogue in DFE-ESMF, as it is a time-independent theory.
Second, the AA, when combined with LR theory to produce practical versions of TDDFT,
creates issues that are not present in DFE-ESMF, regardless of whether ground
state functionals are employed.
Most importantly, the AA prevents TDDFT from incorporating the effects of orbital
relaxations for electrons not involved in the excitation.
Due to its many-electron variational nature, DFE-ESMF explicitly includes these
relaxations, in direct analogy to how ground state KS-DFT variationally relaxes all the
electrons' orbitals.
Thus, although both the AA and the current formulation of DFE-ESMF lead in practice
to the use of ground state functionals for treating excited states, the approximations
being made in these two approaches are distinct.

Finally, it is important to emphasize that state-specific formulations do
come with limitations alongside their advantages.
As for some other excited state specific DFT methods discussed in the next
section, it is not obvious how to arrive at rigorous transition moments
for DFE-ESMF.
Although one could simply define these in terms of the underlying wave function
and the ab initio Hamiltonian, this approach would miss the fact that the
states have been optimized based on a DFT-modified energy expression,
creating a disconnect between the evaluations of energy differences and
transition strengths.
Another issue with state specific methods is that they
do not in general satisfy known sum rules and sum-over-state
expressions. \cite{Parr-Yang}
Thus, although the approach pursued here possesses some formal advantages
when compared to TDDFT, it also suffers from some formal disadvantages.

%Another question that one might have regarding to the usage of ground state 
%functionals to describe excited states is that whether such an 
%approximation is equivalent to AA. We need to stress that although these 
%two approximations have similarities, they are different.
%Firstly, as discussed before, AA fails to account for orbital relaxation effects, while such
%effects are included in DFE-ESMF. Secondly, AA is incapable of describing 
%doubly excited states, while there is no formal difficulty to extend DFE-ESMF to 
%doubly excited states, albeit with a more severe double counting problem since 
%doubly excited states tend to be more complicated. Finally, AA is made in 
%the context of the time-dependent linear response formulation, and the authors
%are not aware of discussions of AA beyond such contexts. Therefore, using 
%ground state functional in DFE-ESMF is fundamentally different from AA in TDDFT.  

%However, DFE-ESMF shares some similarities with the generalized adiabatic connection (GAC)
%excited state method\cite{Gorling1999}. Both DFE-ESMF and GAC use a symmetry-determined 
%linear combination of Slater determinant as wave function to compute kinetic energy, 
%external potential, and exchange energy. In other words, 
%In the initial paper of GAC, the total electron 
%density, rather than the symmetric part of it is used 

%rather, it is an density
%functional extension to ESMF method aiming to capture the missing weak correlations
%in ESMF. 

\subsection{Comparisons to Other State-Specific DFT Methods}
\label{sec:other_methods}

DFE-ESMF is not the first attempt to combine wave function based methods
with density functionals.
For example, multi-reference (MR) DFT methods such as multiconfiguration
Pair-Density Functional Theory (MC-PDFT) \cite{Manni2014, Carlson2015} 
and density matrix renormalization group pair-density functional
theory (DMRG-PDFT) \cite{Sharma2019} also modify a wave function's energy
expression by using an xc energy functional to capture correlation effects.  
A major difference between these methods and DFE-ESMF is that they
target strong correlation in ground states, whereas DFE-ESMF targets
weak correlation in singly excited states.
Another difference is that, because CASSCF and DMRG-SCF  
wave functions already incorporate state-specific orbital relaxations,
the DFT part of their methodology need not address the orbital shapes.
The central feature of DFE-ESMF, on the other hand, is its
excited-state-specific orbital relaxation.
Finally, these MR-DFT approaches use the on-top pair density functional,
\cite{STAROVEROV1999} which is more capable of addressing strong correlation
issues.

With regards to variational DFT methods for excited states, many approaches
distinct from DFE-ESMF already exist.
The $\Delta$ self-consistent field ($\Delta$SCF) approach \cite{Ziegler1997}
relaxes excited state orbitals by using the SCF cycle in an attempt to
converge onto open-shell solutions to KS equations, employing the
maximum overlap method (MOM) to help avoid collapsing back to the ground state
or to lower-lying excited states. \cite{Gilbert2008, Besley2009}
The related restricted open-shell Kohn-Sham (ROKS) method \cite{Kowalczyk2013}
may also collapse to lower excited states, but its enforced open-shell nature
prevents collapse to the ground state and it has shown advantages relative to
$\Delta$SCF for CT excitations' singlet-triplet splittings. \cite{Hait2016}
Finally, ensemble DFT in the form of REKS and SA-REKS optimizes excited
state orbitals in a state-averaged manner, trading some state-specificity
in return for a further reduction in the risk of variational collapse.
In contrast to these approaches, the DFE-ESMF approach makes direct use
of an excited state variational principle.
Although this does not rigorously guarantee that the correct stationary
point will be found (all of these variational methods are nonlinear
minimizations with at least some starting point dependence, after all)
the global minimum of the variational principle it employs is the
desired excited state, offering a strong formal advantage in the
effort to avoid collapse to lower states.
In our preliminary explorations, we have yet to encounter a case where
the optimization does not converge to the stationary point corresponding
to the targeted excited state, even in cases where $\Delta$SCF
encounters variational collapse.
It is also worth noting that, although the multi-CSF version of DFE-ESMF
comes with double counting concerns, it can at least be applied to states
that strongly mix two or more excitation components, while
$\Delta$SCF, ROKS, and REKS all assume that excitations are
single-component in nature.

%\subsection{Comparison to Constrained DFT}
%\label{sec:cdft}

The constrained DFT (CDFT) method \cite{Kaduk2012} 
represents another route towards excited state orbital relaxation that is
especially relevant for long range CT, where it is straightforward
to impose physically motivated density constraints in cases where the
donor and acceptor can be clearly identified.
As shown by numerous applications, CDFT can provide accurate estimates
of excitation energies, \cite{Wu2006} coupling elements, \cite{Wu_jcp2006}
forces, \cite{Wu_jpca2006} and diabatic surfaces. \cite{kaduk2010}
A particularly strong parallel with DFE-EMSF can be seen in long range
CT, where single-CSF DFE-ESMF is expected to be equivalent to CDFT
in the limit of complete donor-acceptor separation
(see for example Figure S1).
To understand this equivalence, consider that both methods will
move an electron from the donor's HOMO to the acceptor's LUMO and then
make their energy expression stationary with respect to orbital rotations.
As the WFCE term in DFE-ESMF vanishes in the limit of long range CT,
the two methods will have the same energy expression in this case
and so will produce the same results.
In shorter-ranged CT where donor and acceptor are less well defined,
DFE-ESMF has the formal advantage of not having to impose a user-specified
charge constraint, and so can in principle predict the distribution of
the particle and hole rather than having it imposed from some external source.
DFE-ESMF also avoids having to worry about the ambiguities
inherent to assigning formal atomic charges and the difficulties these create.
\cite{Kaduk2012}

\section{Results}
\label{sec:results}

\subsection{Computational Details}
\label{sec:compdetail}

To assess the performance of DFE-ESMF and to compare
to existing excited state DFT methodologies,
we have carried out tests in the following
atomic and molecular excitations:

1) singlet and triplet $n\rightarrow\sigma^{\ast}$ in H$_2$O.

2) singlet and triplet $n\rightarrow\pi^{\ast}$ and singlet $\pi\rightarrow\pi^{\ast}$ in CH$_2$O.

3) singlet and triplet $\sigma\rightarrow\sigma^{\ast}$ in LiH.

4) singlet $\pi\rightarrow\pi^{\ast}$ in CO. 

5) singlet He 1s$\rightarrow$Be 2p in He-Be dimer.

6) singlet NH$_3$ 2pz$\rightarrow$F$_2$ 2pz in  NH$_3$-F$_2$ dimer.

7) singlet 2s$\rightarrow$3s and 2p$\rightarrow$3p in the Ne atom. 

\noindent
All of the DFE-ESMF results are obtained via our own pilot code, which extracts 
one- and two-electron integrals from PySCF.
\cite{Sun2018}
The Lebdev-Laikov 
grid\cite{lebdev1999} is used to perform the numerical integration to compute the xc energy. 
The TDDFT, CIS, ROKS, and $\Delta$SCF DFT results were
obtained from QChem. \cite{qchem}
Equation-of-Motion Coupled Cluster with singles and doubles (EOM-CCSD) results were
computed by MOLPRO. \cite{Knowles2011} It is also worth noting that the 
implementation of ROKS in QChem is limited to only HOMO$\rightarrow$LUMO 
excitations. 
In the current study, the CSF expansions in both the
single-CSF formalism and multiple-CSF formalism are selected by the CI vector of TDDFT 
using the same xc functional. We choose a large threshold of $\epsilon=0.2$
for CSF truncation, and switch to the multi-CSF
formalism of DFE-ESMF when there are more than one CSF left after truncation. 

For DFE-ESMF, we employ three xc functionals: LDA, the Becke3-Lee-Yang-Parr
functional (B3LYP),\cite{Lee1988,Becke1988a,Becke1998} and the
Becke-Half-Half functional (BHHLYP),\cite{Becke1993} 
which have 0\%, 20\%, and 50\% wave function exchange fractions, respectively.
We expect results to be somewhat sensitive to this fraction,
as ground and excited states have different amounts
of open shell character and thus are likely to suffer from differing
degrees of self-interaction error.
As existing functionals have mostly been optimized for closed shell
ground states, it would not be surprising if a higher than usual
wave function exchange fraction was necessary to
create a fair playing field for the open-shell state.
Note that our excitation energies come from energy differences between
DFE-ESMF (for the excited state) and KS-DFT (for the ground state) in
which both have used the same xc functional.

%In H$_2$O the H-O-H bond angle is chosen to be 104.5$^\circ$ and the O-H bond
%length is 0.96$\mathring{A}$. In CH$_2$O the H-C-H bond angle is 116$^\circ$, 
%and the C-H and C-O bond length are 1.11$\mathring{A}$ and 1.21$\mathring{A}$
%respectively. In LiH the bond length is 1.6$\mathring{A}$, and the He-Be separation is 3.5$\mathring{A}$.
%The geometry 
%of the NH$_3$-F$_2$ dimer is shown in Figure \ref{fig:nh3_f2_geo}.
For basis sets, we employed the cc-pVDZ basis \cite{Dunning1988} for H$_2$O, CH$_2$O, LiH, and CO, the cc-pVTZ basis \cite{Kendall1992}
for the He-Be dimer, the aug-cc-pVTZ basis \cite{Kendall1992} for Ne, and the 6-31G basis \cite{Hehre1972} for the NH$_3$-F$_2$ dimer. 
Results will be presented in terms of excitation energy errors relative to EOM-CCSD. 
The molecular geometries and absolute values of excitation energies can be found 
in the Appendix.

%\begin{figure}
%\centering
%\includegraphics[width=8.5cm,angle=0,scale=0.7]{new_NH3_F2.png}
%\caption{Geometry of the NH$_3$-F$_2$ dimer. The N-H bond length and 
%H-N-H bond angle is 1.02$\mathring{A}$ and 106.2$^\circ$. The F$_2$ bond length
%is 1.43$\mathring{A}$. 
%%{\color{red} Need to give FF and NH bond lengths.}
%        }
%\label{fig:nh3_f2_geo}
%\end{figure}

\subsection{Excited State Dipole Shifts}
\label{sec:dipole}

Before presenting the results of DFE-ESMF, we first categorize the excited states 
into CT and non-CT types by computing the difference between the ground and 
excited state dipole moment.
In atomic units, the dipole moment is
\begin{equation}
    \label{eqn:dip_moment}
    \vec{\mu}=
    \sum_{A}{Z_A\textbf{R}_A}
    - \int{\textbf{r} \hspace{.4mm} n(\textbf{r}) \hspace{.2mm} d\textbf{r}}
\end{equation}
in which $Z_A$ and $\textbf{R}_A$ are the charge and position of the $A$th nuclei.
For simplicity, the electron density for excited states is estimated using Equation \ref{eqn:exden}
using ground state KS orbitals without any relaxation.
The dipole moment 
difference (|$\Delta \vec{\mu}$|) between the ground and excited state yields 
information about the electron charge distribution between these two states. 

\begin{figure}
\centering
\includegraphics[width=8.5cm,angle=0]{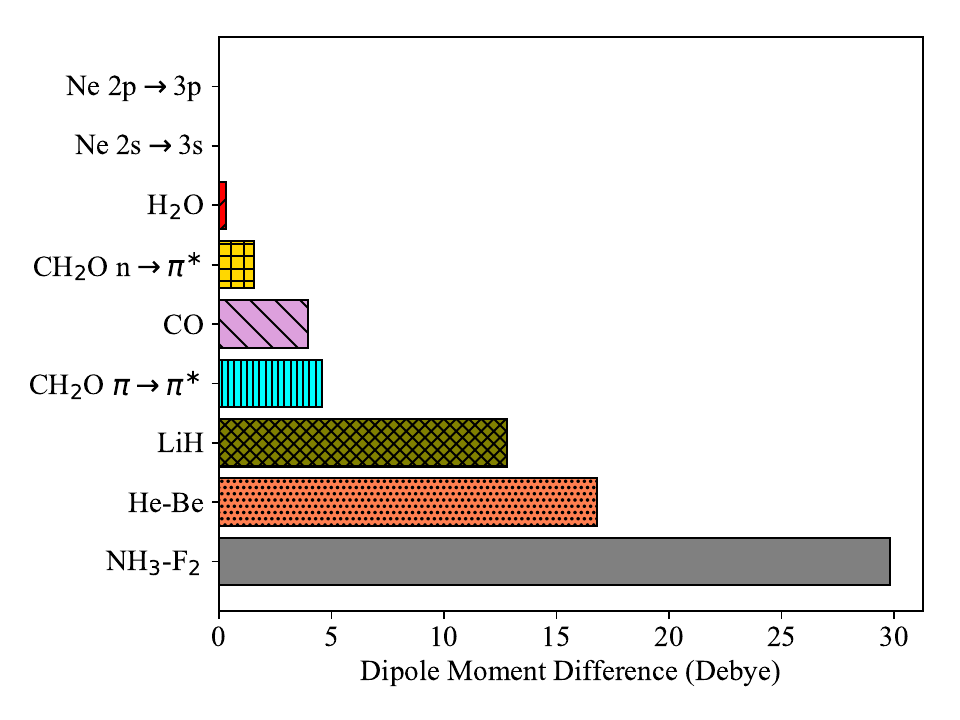}
\caption{The norms of the dipole moment differences between ground and excited states. 
The BHHLYP functional is used in all dipole calculations. 
%{\color{red} Font size in this figure needs to be at least as large as the text in this caption.}
        }
\label{fig:dip_diff}
\end{figure}

The computed |$\Delta \vec{\mu}$|s are shown in Figure \ref{fig:dip_diff}. 
For Ne, H$_2$O, and the $n\rightarrow \pi^{*}$ state in CH$_2$O,
the |$\Delta \vec{\mu}$| are fairly small, indicating
that the charge distributions are similar for the ground and excited states. 
These states can thus be viewed as purely Rydberg (Ne) and 
valence (H$_2$O, CH$_2$O) excited states with little charge deformation.
As expected, the long-range CT excited states \cite{Zhao2006,Gritsenko2004}
of He-Be and NH$_3$-F$_2$ have a significantly larger |$\Delta \vec{\mu}$|.  
Note that LiH also sees significant charge deformations,
which are a consequence of its partially ionic nature in which the bonding
and anti-bonding orbitals are shifted towards opposite ends of the molecule. 
%CO and $\pi\rightarrow\pi^{\ast}$ CH$_2$O also have sizable but not too large 
%|$\Delta \vec{\mu}$|, this suggests that these two valence excited states are also
%accompanied with some amount of charge deformation.  
%In summary, we categorize H$_2$O and CH$_2$O as valence 
%excitations, Ne as Rydberg excitation, 
%and LiH, He-Be, and NH$_3$-F$_2$ as CT excitations. 

\subsection{Single-CSF Excited States}
\label{sec:singleCSFResults}

Let us begin with an analysis of how the theory performs for excited states
dominated by a single CSF, looking at different types of excitations
within this category.
We will look first at valence excitations, followed by CT states
and finally Rydberg excitations.
These cases covered, we will then consider states that contain
a superposition of multiple CSFs.

\subsubsection{Valence Excitations}
\label{sec:valence}

%We explore the VES-DFT method with three different xc functionals 
%that include different mixing of the exact exchange in this study: the local density 
%approximation (LDA, 0\% of exact exchange), the Becke3-Lee-Yang-Parr functional\cite{Lee1988,Becke1988a,Becke1998} (B3LYP, 20\%
%of exact exchange), and the Becke-Half-Half\cite{Becke1993} (BHHLYP, 50\% of exact exchange). 

\begin{figure}
\centering
    \includegraphics[width=8.6cm,angle=0]{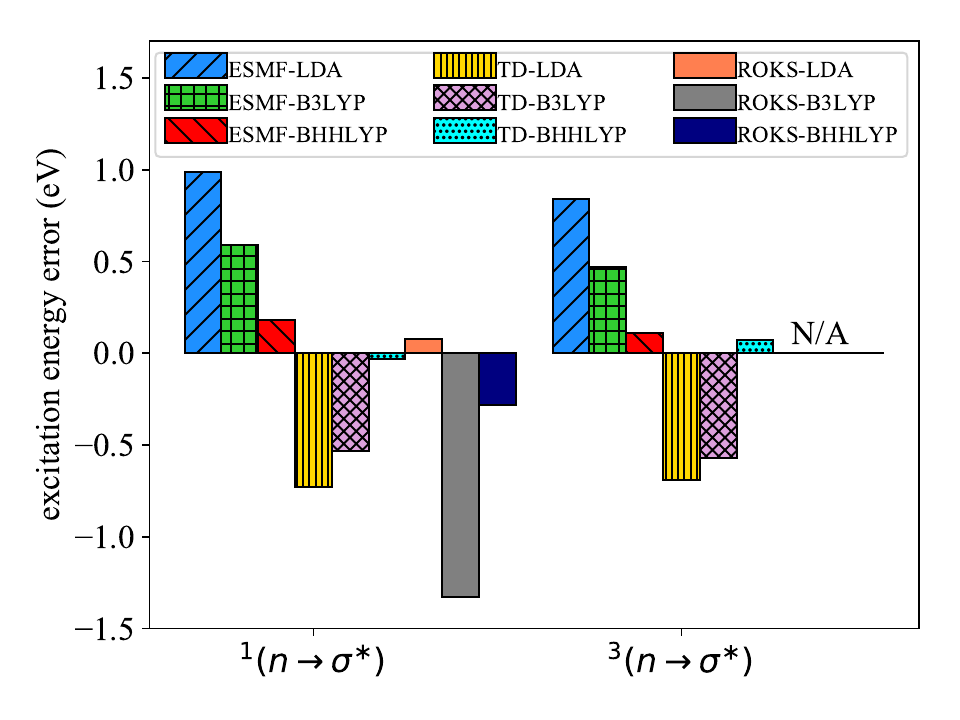}
\caption{The excitation energy error of singlet (left) and triplet (right)
         $n\rightarrow\sigma^{\ast}$ excited states in H$_2$O 
         compared to EOM-CCSD results. 
        }
\label{fig:h2o}
\end{figure}

\begin{figure}
\centering
\includegraphics[width=8.6cm,angle=0]{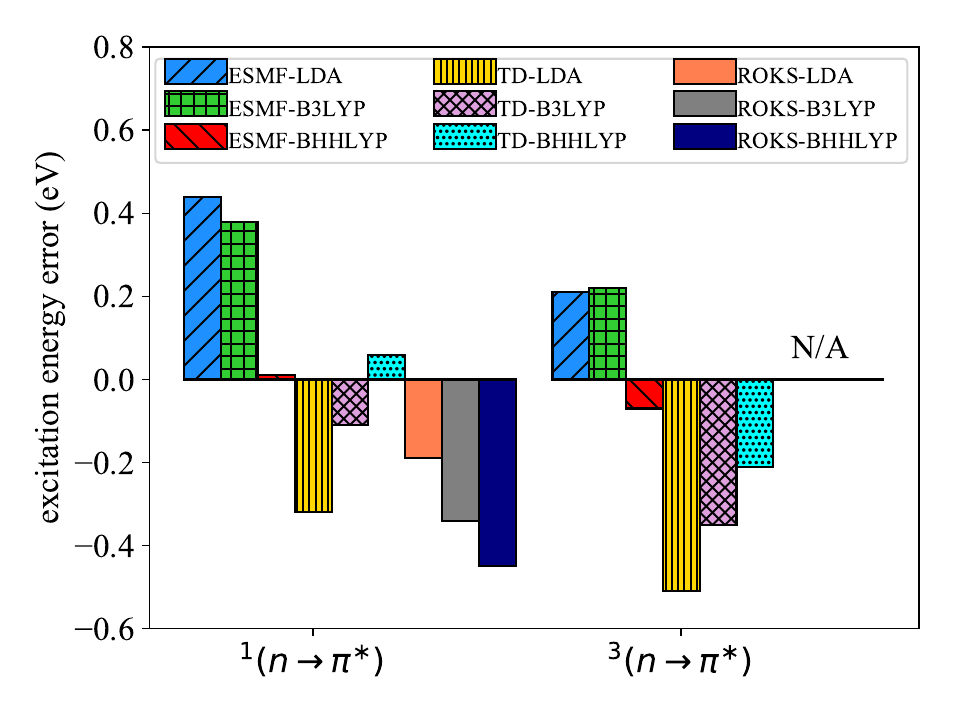}
\caption{The excitation energy error of singlet (left) and triplet (right)
         $n\rightarrow\pi^{\ast}$ excited states in CH$_2$O
         compared to EOM-CCSD results.
        }
\label{fig:ch2o}
\end{figure}

Excitation energy errors relative to EOM-CCSD for H$_2$O
and CH$_2$O are shown in
Figures \ref{fig:h2o} and \ref{fig:ch2o}, respectively.
Compared to the CT examples in the next section, TDDFT
performs relatively well for these valence excitations,
with B3LYP and especially BHHLYP providing excitation
energies within about half an eV of the reference and
LDA performing only a little worse. 
These relatively good TDDFT results are not particularly
surprising in light of the fact that the charge density
deformations are small, and so the lack of orbital
relaxation due to the AA is not especially concerning.
In fact, we have explicitly analyzed the importance of orbital relaxation by evaluating the Frobenius norm of the DFE-ESMF
orbital rotation matrix $\bm{X}$.
Averaging over the three values from the three different
functionals tested, we find $||\bm{X}||$ to be
0.16 and 0.15 for H$_2$O and CH$_2$O, respectively,
which is smaller than in the CT examples we will see below.

The basic trend in DFE-ESMF accuracies for different
functionals follows that of TDDFT in these excitations,
with BHHLYP giving the most accurate predictions,
followed by B3LYP and then LDA.
As we will see, the accuracy ordering for different functionals
in DFE-ESMF plays out is the same way in most of our test systems,
with BHHLYP's high fraction of wave function exchange outperforming
the other two functionals in valence, CT, and Rydberg states.
Our understanding of this trend is that
a larger fraction of wave function exchange is most likely helping
to balance self-interaction errors in the ground and
excited states.
We expect that these errors are larger in the excited states due
to their open-shell nature, and so a higher fraction of
wave function exchange than is typically used in ground state
models appears to be helpful for balancing these errors
between the ground and excited states.
We should emphasize that in all of our DFE-ESMF results,
energy differences were evaluated based on the same
functional for both ground and excited states, but
using the density and wave function exchange
definition for the state in question
(see discussion surrounding Eq.\ (\ref{eqn:exDFT_eng})).
Although this preliminary test of four single-CSF valence
excitations is far from exhaustive or systematic,
the efficacy of DFE-ESMF/BHHLYP in these cases
provides an encouraging proof of principle.

For singlet excited states, we also compare to the predictions of
ROKS, which, like DFE-ESMF, provides excited state orbital relaxation.
The most striking difference between the ROKS results and those of TDDFT
and DFE-ESMF is that they do not follow the same trend with respect to
the fraction of wave function exchange.
Indeed, the accuracy ordering differs in H$_2$O and CH$_2$O, with LDA
seeming to offer the best average ROKS performance and with
ROKS/B3LYP delivering a surprisingly large error in H$_2$O.
Note that we have not carried out ROKS comparisons in the triplet
states because QChem currently only implements ROKS for singlet states.

%\begin{figure}
%\centering
%\includegraphics[width=8.5cm,angle=0]{hf_error.pdf}
%\caption{The excitation energy errors relative to EOM-CCSD for the HF
%         $\pi\rightarrow\pi^{\ast}$ singlet excitation. 
%        }
%\label{fig:hf}
%\end{figure}

\subsubsection{Charge-Transfer Excitations}
\label{sec:ct}

Although DFE-ESMF and TDDFT provide similar accuracies in
the simple valence excitations discussed above, the story
is very different for CT excitations.
Before looking in detail at the numerical CT examples, it
is worth considering two important potential sources of
error TDDFT faces in CT contexts, which we will refer to
as the EA/IP imbalance and the orbital relaxation error.
To see these clearly, consider a simple CT excitation
consisting of a single $i\rightarrow a$ transition, in
which case the TDDFT excitation energy is given by
\cite{Baerends2013}
\begin{equation}
    \label{eqn:tddft_ext}
    \Delta E(i\rightarrow a)=\epsilon_a^{KS}-\epsilon_i^{KS}+\left<ia|f_{xc}|ia\right>
\end{equation} 
in which $\epsilon_i^{KS}$ and $\epsilon_a^{KS}$ are the
ground state KS orbital energies of the donor and acceptor orbitals, respectively.
% Since the orbital relaxation effect is small, the remaining error in both 
% theories is then due to the usage of approximate xc functionals. However, 
% as we shall discuss, the situation becomes completely different for
% CT excited states, in which TDDFT with AA yields substantial amount of 
% errors and going beyond the frozen orbital effects of AA becomes necessary. 
%The well-known issue with the
%lack of a Koopman's theorem As is well known (see also Equation \ref{eqn:ksdft_sc}),
This equation has been used extensively\cite{Dreuw2003,Dreuw2005,Maitra2016,Maitra2017} to analyze the 
TDDFT's failure in CT excited states and we refer readers to those references for 
details.
In a nutshell, because $\epsilon_a^{KS}$ does not correspond to the EA
of the acceptor (it undercounts the new repulsions created by the extra electron),
the orbital energy difference in this equation tends to severely underestimate
CT excitation energies.
In principle this should be repaired by the xc term, but the error is often
much larger than existing functionals, even RSHs, can correct for.

While the EA/IP imbalance is a significant concern, it is typically offset
in practice by the fact that TDDFT works with unrelaxed orbitals.
It is well known in electronic structure theory that the omission of 
orbital relaxation effects tends to raise the energy of the excited state. 
%As orbital relaxations in the excited state are expected to lower the
%energy of the acceptor orbital and raise the energy of the donor orbital,
%the orbital relaxation error arising from the lack of these effects
%%in the orbital energy difference of Equation \ref{eqn:tddft_ext}
%will work to push excitation energies upwards.
%%inclusion of orbital relaxations would be
%%expected to push towards lower excitation energies.
%%The lack of such relaxations (i.e.\ the ORE)
%%will therefore push excitation energies to be too high.
Ideally, the third term in Equation \ref{eqn:tddft_ext} would
eliminate both orbital relaxation issues and the EA/IP imbalance,
but even when the third term is zero, these two errors do at least
work to cancel each other because they push in opposite directions.
However, the EA/IP imbalance in long-range CT is often much too large
for orbital relaxation errors to counteract, resulting in TDDFT excitation
energies that are much too low as in the NH$_3\rightarrow$F$_2$
and He$\rightarrow$Be transitions shown below.
However, as one moves to increasingly shorter range CT with
correspondingly larger overlaps between the donor and acceptor,
the EA/IP imbalance becomes less and less of an error and more and more
a positive feature of the TDDFT formalism.
Indeed, in the valence excitation limit, the difference in how DFT
accounts for electron-electron repulsion energy in the occupied
and virtual orbitals increases the accuracy of using orbital energy
differences as excitation energy estimates, because in this limit
the ``donor'' and ``acceptor'' are one and the same.
One can imagine that for very short-ranged CT, any small remaining
errors from the EA/IP imbalance could cancel with orbital relaxation
errors precisely enough for the exchange correlation term to
clean up the details.
Such an effect seems to be at work in LiH, to which
we now turn our attention, which despite having a substantial dipole
change and thus CT is nonetheless treated well by TDDFT.

\paragraph{LiH}
\label{sec:lih}

%% we could see that in KS-DFT, both
%occupied and virtual orbitals feel the same potential created by 
%repulsion of $N$ electrons and attraction of the ``xc-hole'' 
%$\upsilon_{xc}(\textbf{r})$, which integrates to -1 electron.  
%Therefore, the excited electron in virtual orbital ``sees''
%the same $N-1$ electrons as do the electrons in ground state\cite{Baerends2013}. 
%Owing to this, the orbital energy difference ($\epsilon_a^{KS}-\epsilon_i^{KS}$) 
%is already a good estimation 
%to the excitation energy, and the $\left<ia|f_{xc}|ia\right>$ term
%only provides small corrections. 

\begin{figure}
\centering
\includegraphics[width=8.5cm,angle=0]{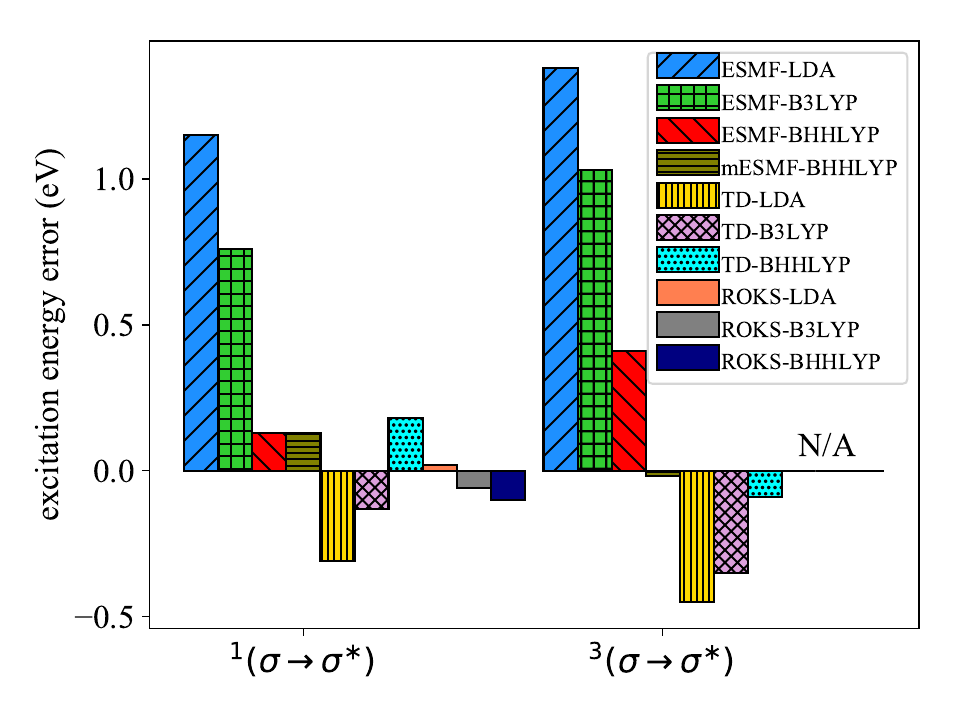}
\caption{The excitation energy error of singlet (left) and triplet (right)
         $\sigma\rightarrow\sigma^{\ast}$ excited states in LiH
         compared to EOM-CCSD results.
        }
\label{fig:lih}
\end{figure}

Figure \ref{fig:lih} shows excitation energy errors for the lowest singlet
and triplet excitations in LiH.
For both of these states, the balancing process between EA/IP issues and
missing orbital relaxations appears to work in TDDFT's favor, especially
in the case of the BHHLYP functional.
To check that such a trade off really does appear to be at work here,
we again evaluated $||\bm{X}||$ as a measure of orbital relaxation
importance and found it to be 0.41, significantly higher than
for H$_2$O or CH$_2$O.
As in the valence states of those molecules, BHHLYP is also very effective
in DFE-ESMF's single-CSF formalism for LiH's singlet excitation.
For the triplet, however, the single-CSF formalism shows relatively 
poor accuracy regardless of functional, and indeed this state has more than one
CSF above our $\epsilon=0.2$ threshold in TDDFT.
When we include both of the CSFs whose coefficients breach this threshold
via the multi-CSF approach, the DFE-ESMF/BHHLYP result improves from an error
above 0.4 eV to an error of just -0.02 eV relative to EOM-CCSD.
Note that the multi-CSF approach has no effect on the singlet state,
as in that case only the primary CSF was above the threshold.

\paragraph{NH$_3$ to F$_2$}
\label{sec:nh3f2}

\begin{figure}
\centering
\includegraphics[width=8.5cm,angle=0]{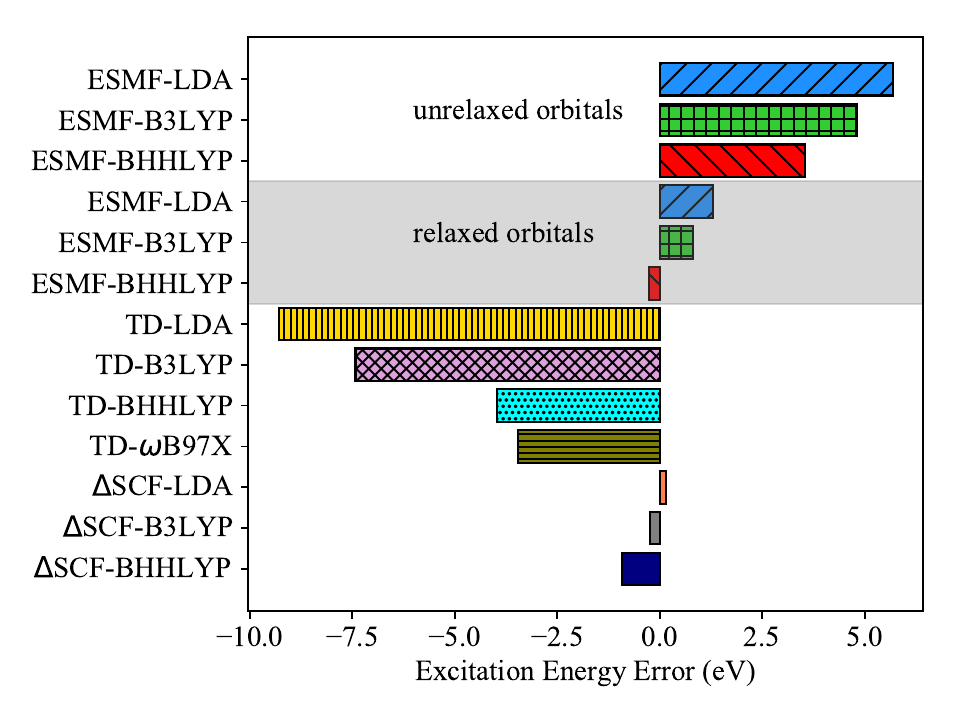}
\caption{Comparison of excitation energy errors relative to EOM-CCSD
         for the NH$_3$ 2pz $\rightarrow$F$_2$ 2pz CT excitation at
         an intermolecular separation of 6 $\mathring{A}$.
         For DFE-ESMF, we show the results both before and after the
         orbital relaxation is performed.  
         %The NH$_3$-F$_2$ . 
        }
\label{fig:nh3_f2}
\end{figure}

We now turn our attention to the first of two long-range CT excitations:
the NH$_3$-F$_2$ dimer shown in Figure S1.
%As shown in Figure \ref{fig:nh3_f2_geo}, the 
%lone pair of the NH$_3$ and the F$_2$ bond are aligned along the z-axis and the separation
%between these two molecules is set to be 6 $\mathring{A}$.
In Figure \ref{fig:nh3_f2}, we see that, after excited-state orbital relaxation
via the minimization of Eq.\ (\ref{eqn:esmf_approx_L}), DFE-ESMF is substantially
more accurate than TDDFT regardless of the functionals chosen.
Even when comparing ESMF-LDA against TDDFT with the $\omega$B97X RSH, the variational
approach makes an excitation energy error roughly half as large.
Using BHHLYP, which continues to outperform the others for DFE-ESMF, the variational
approach achieves an excitation energy error of just 0.26 eV, compared to
multi-eV errors for TDDFT when using either $\omega$B97X or BHHLYP. 

Although this excited state is predominantly a HOMO$\rightarrow$LUMO excitation,
we find that ROKS consistently collapses to a lower excited state (see the Appendix
for absolute excitation energies). 
Hence, we compare our results to MOM $\Delta$SCF instead.
%The broken of spin symmetry
%in $\Delta$SCF DFT would only slightly affect the energy in this case, due to 
%the large separation between donor and acceptor.
We found that $\Delta$SCF DFT 
yields similarly accurate predictions as DFE-ESMF. This is as expected since 
$\Delta$SCF DFT are known to perform well in long-range CT excited states, if the 
states are dominated by one CSF. 
%The similar accuracy of DFE-ESMF and $\Delta$SCF DFT is as expected,  
%since in they both  

It would appear that TDDFT's error cancellation between its EA/IP imbalance and
its lack of orbital relaxations breaks down here, with the magnitude of the former
overwhelming that of the latter.
We can verify this picture in two ways:  first, with the DFE-ESMF approach, and second,
by looking at ground state KS-DFT IP-EA estimates at very long range.
Start with DFE-ESMF.
At the top of Figure \ref{fig:nh3_f2}, we show the excitation energies (i.e.\ the energy
differences between Eq.\ (\ref{eqn:exDFT_eng}) and the ground state KS energy)
\textit{before} the DFE-ESMF orbital optimization has been carried out, meaning that
the excited state DFE-ESMF energy is being evaluated using the ground state KS orbitals.
This excitation energy is thus the difference between two many-electron
DFT energies (one DFE-ESMF and one KS-DFT) and so does not suffer from the EA/IP
imbalance. 
%that arises in the difference between KS single-particle orbital energies.
While the EA/IP issue has thus been removed, orbital relaxation effects have yet to
be included, and as expected the excitation energies are now too large.
When we then relax the orbitals ($||\bm{X}||=0.21$), we see in the middle of Figure \ref{fig:nh3_f2}
that the predicted excitation energies decrease to more accurate values.
Thus, by looking step-wise at how DFE-ESMF changes the energy from TDDFT, we can watch
the staged removal of first the EA/IP imbalance and then the fixed-orbital error.
This process appears to confirm the idea that TDDFT suffers from both, and that in
long-range CT the EA/IP part dominates, leading TDDFT to underestimate the excitation energy.

\begin{figure}
\centering
\includegraphics[width=8.5cm,angle=0]{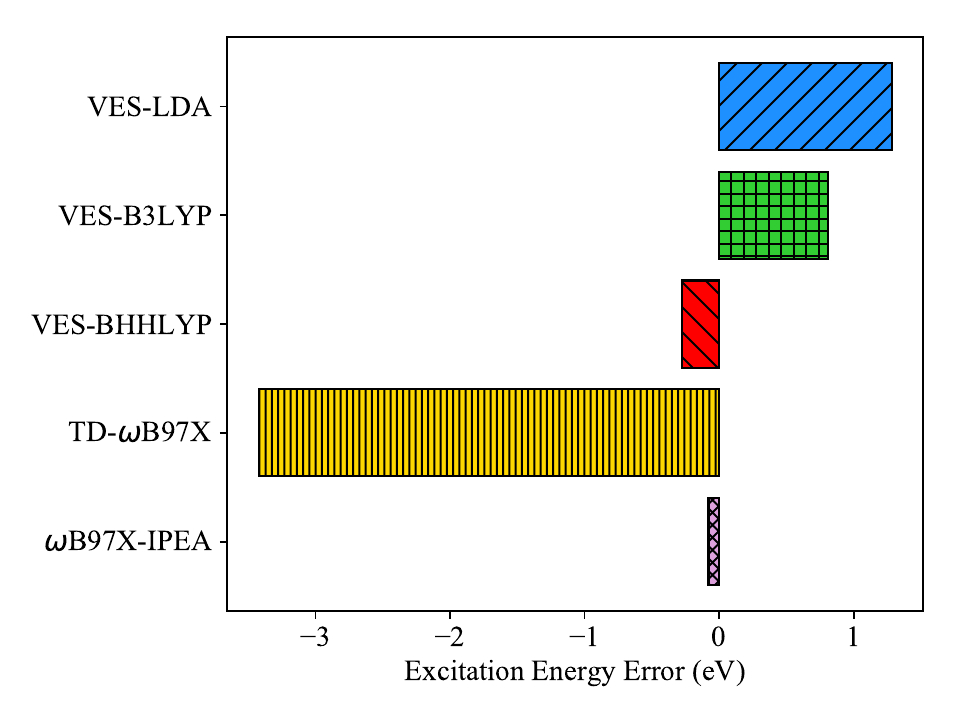}
\caption{Excitation energy errors relative to EOM-CCSD
         for the NH$_3$ 2pz $\rightarrow$F$_2$ 2pz CT
         at a 120 $\mathring{A}$ intermolecular separation. 
         $\omega$B97X-IPEA refers to the difference between
         (a) the sum of ground state KS energies for the
         donor cation and acceptor anion and (b) the
         ground state KS energy of the neutral ground states.
        }
\label{fig:nh3_f2_inf_sep}
\end{figure}

We can corroborate this view by moving the molecules to a very large distance
and comparing DFE-ESMF, TDDFT, and a simple difference of ground state KS
energies between the cation, anion, and neutral species that provides
a many-electron evaluation of the IP and EA.
At very long distance, Figure \ref{fig:nh3_f2_inf_sep} shows that the
performance of DFE-ESMF and TDDFT is quite similar to what we saw at the
shorter separation.
At the bottom of the figure, we see that if we simply perform four
single-molecule ground state KS calculations for the donor cation,
acceptor anion, and the two neutral species, the resulting difference
between the IP and EA is a very accurate predictor of the charge
transfer energy, as we would expect from previous work on CDFT.
Thus, if both the IP/EA imbalance born of single-particle orbital
energy differences and the orbital relaxation errors are removed
via this ground state KS approach or by CDFT, accuracy is restored.
The key point is that, unlike these approaches, the DFE-ESMF formalism should allow both of these issues to be
addressed even in systems where clear-cut foreknowledge distinguishing
between the donor and acceptor is not available, and without
having to worry about the spin-symmetry breaking inherent
to $\Delta$SCF.

\paragraph{He to Be}
\label{sec:hebe}

\begin{figure}
\centering
\includegraphics[width=8.5cm,angle=0,scale=1.0]{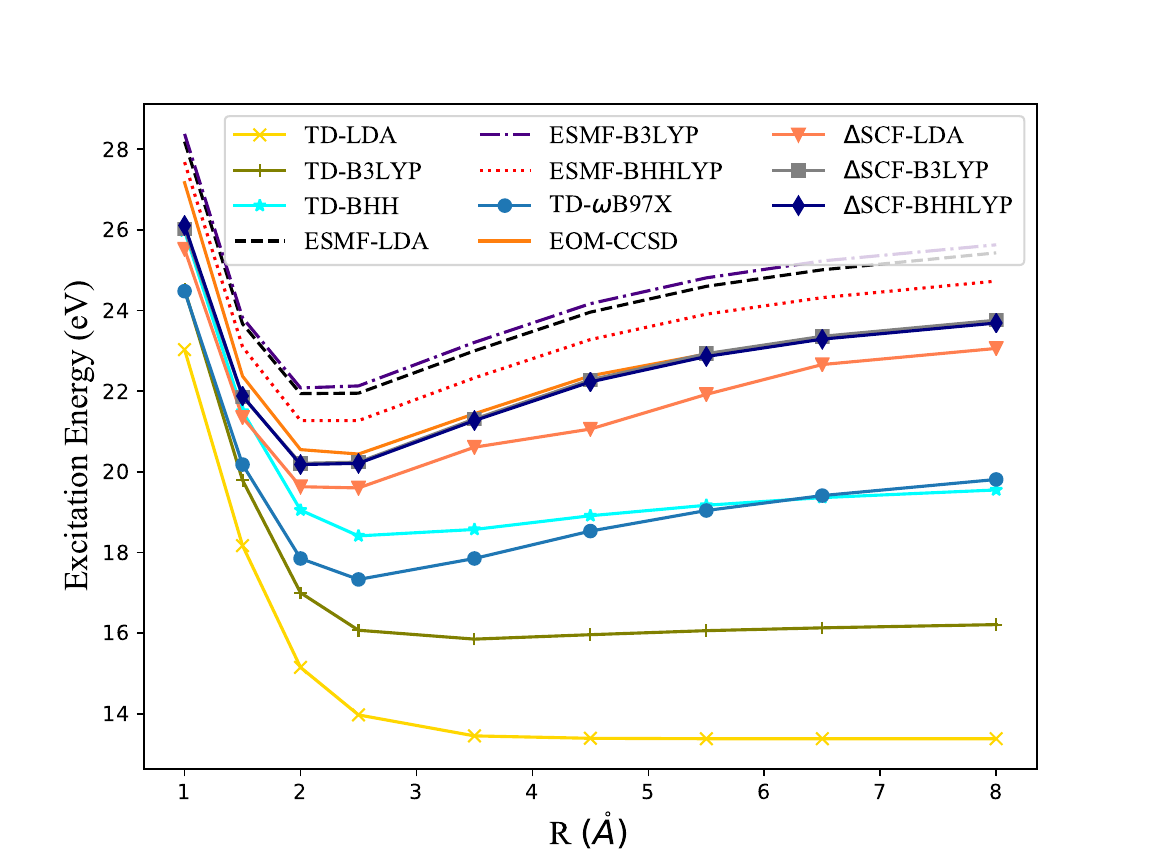}
\caption{Comparison of He 1s$\rightarrow$Be 2p CT excitation energy as a function 
         of $R$(He-Be) between DFE-ESMF, TDDFT, $\Delta$SCF, and EOM-CCSD. 
        }
\label{fig:he_be}
\end{figure}

\begin{figure}
\centering
\includegraphics[width=8.5cm,angle=0]{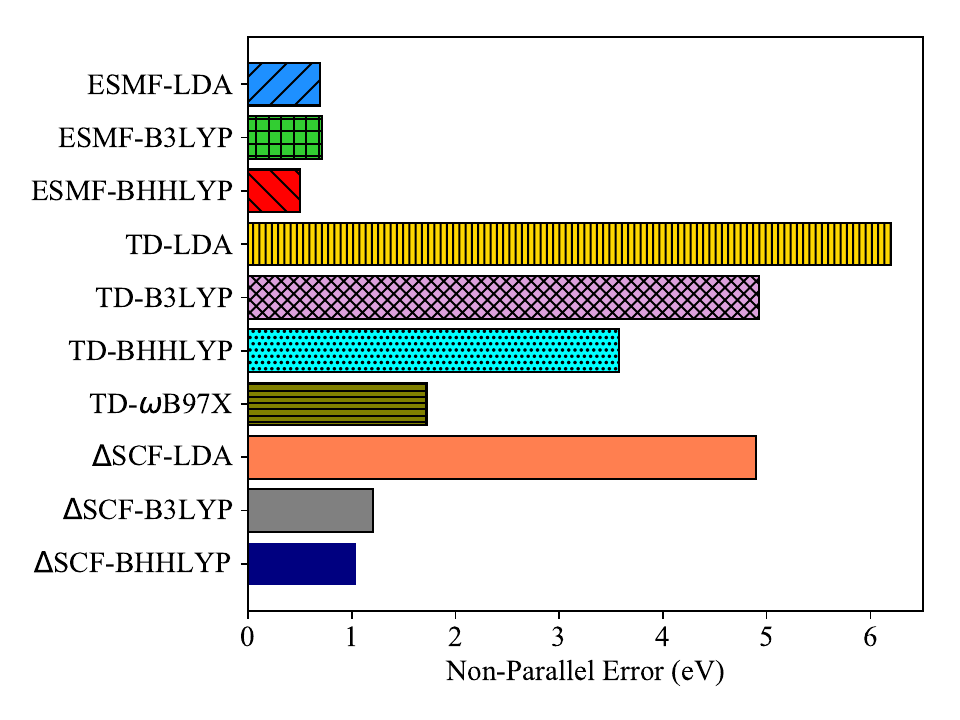}
\caption{Comparison of non-parallelity errors in the He 1s$\rightarrow$Be 2p CT
          excitation energy. 
        }
\label{fig:he_be_non_parallel}
\end{figure}

In our second long-range CT example, we investigate the excitation
from the He 1s orbital to the Be 2pz orbital as a function of
the distance between the atoms.
In Figure \ref{fig:he_be}, we see the familiar failure of
local functionals and simple hybrids to predict the correct
$1/R$ trend in the excitation energy.
While this problem is repaired by the use of a RSH, we see that
absolute accuracy is still poor, at least for the specific
$\omega$B97X functional we tested here.
As in the previous example, DFE-ESMF out-performs the
accuracy of the RSH regardless of which functional it is
paired with while also correctly capturing the $1/R$
behavior.
The advantage of DFE-ESMF becomes especially clear when looking
at the non-parallelity error (NPE) plotted in Figure 
\ref{fig:he_be_non_parallel}, which is defined as the
difference between the largest and smallest errors
relative to EOM-CCSD across the distance coordinate.
DFE-ESMF with LDA, B3LYP, and BHHLYP all produce NPEs below 1eV, as compared
to an NPE of almost 2 eV for TDDFT with the $\omega$B97X functional.

While $\Delta$SCF produces a visibly not-smooth curve when
when paired with the LDA functional
(possibly due to variational collapse issues),
its performance with B3LYP and BHHLYP is quite good.
In the long-range limit these potential curves overlap that
of EOM-CCSD, although accuracy is a bit lower at
shorter ranges where the broken spin symmetry is expected
to matter more.
Consequently, the NPEs of $\Delta$SCF with the hybrid functionals
are a bit larger than those of DFE-ESMF, but a major
improvement over TDDFT.

\subsubsection{Rydberg Excitations}
\label{sec:rydberg}

%{\color{red}
%Why didn't you use diffuse functions in the basis set?
%Without doing at least aug-cc-pVTZ, I'm not sure we can
%make firm conclusions about whether or not our orbital
%rotations actually help via Troy's arguments about
%the xc potential from the excited state density.
%}
Unlike in CT excitations, it is not obvious that the ability of DFE-ESMF
to address the EA/IP imbalance and orbital relaxations will be of great
benefit in the context of Rydberg excitations.
In these states, the challenge faced by TDDFT arises primarily from the
failure of practical xc functionals to produce a potential that decays as
$1/r$ at large distances, which is not the same as the failure of error
cancellation that causes trouble in the CT case.
However, work by Van Voorhis
%Rydberg excitations represent another type of excited states that can not be well 
%described by TDDFT. As discussed before, Rydberg states are defined as excited states
%in which one electron has been excited to a high principle quantum number orbital.
%As discussed before, the failure TDDFT for Rydberg excited states is due to
%the wrong asymptotic behavior of the ground state xc potential. The 
%exact xc potential should decay as $1/r$ at large distances, while commonly used 
%functionals decay exponentially.
%At a first glance, this problem does not 
%stem from the lack of orbital relaxation or IOPE and VES-DFT
%is of little use.
%However, Troy Van Voorhis and coworkers
\cite{Cheng2008}
has shown that although the ground state xc potential 
has little resemblance to the exact potential, the xc potential associated
with an excited state density can behave much more sensibly at long distance.
%is actually close to the exact value.
Thus, it is interesting to ask whether DFE-ESMF's inherently excited state nature
and its ability to relax the orbitals in an excited-state-specific manner
may in practice lead to improvements for Rydberg states.
%Like the CT excitations, Rydberg excitations are also accompanied by
%drastic density deformation from the ground state, owing to the difference in shape 
%and spatial occupancy between the occupied and virtual orbitals. Therefore 
%we expect that orbital 
%relaxation is also needed to obtain accurate predictions
%to the excitation energy. 

\begin{figure}
\centering
\includegraphics[width=8.5cm,angle=0]{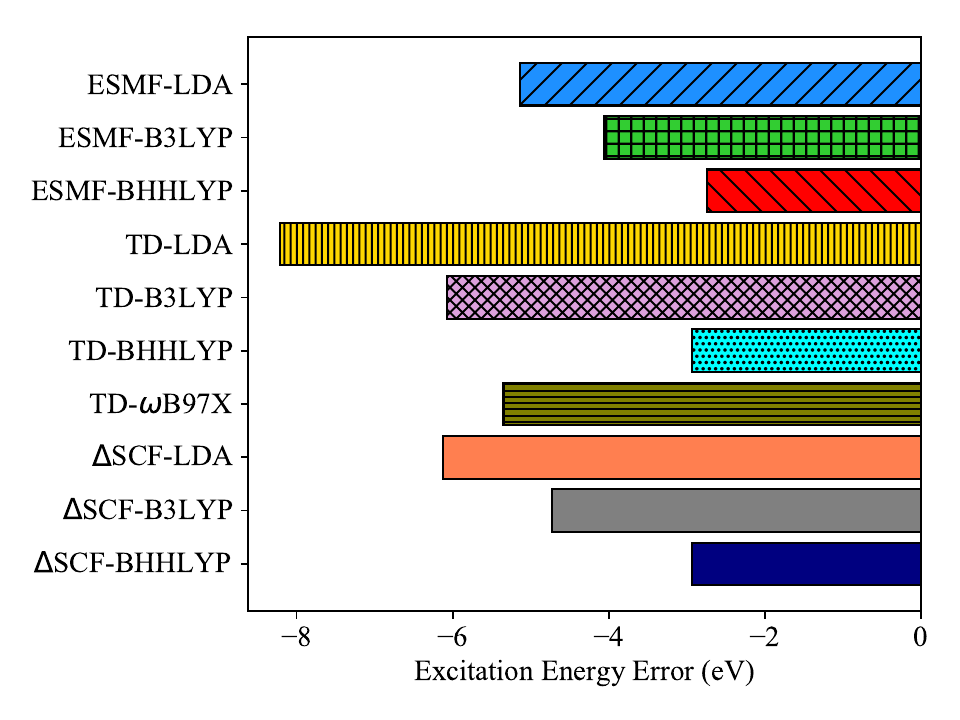}
\caption{Comparison of excitation energy error of the Ne 2s
         $\rightarrow$ 3s transition. 
        }
\label{fig:ne}
\end{figure}

As an initial probe of this question, we have studied the
2s$\rightarrow$3s excitation in the Ne atom.
The excitation energy error relative to EOM-CCSD is plotted in Figure \ref{fig:ne}. As expected, TDDFT 
drastically underestimates the excitation energy by as large as 8.21eV using LDA and 
2.94eV using BHHLYP. Notably, although our DFE-ESMF method is able to reduce the 
error of TDDFT by some amount, it is still very far from being quantitatively accurate. 
The most accurate functional in DFE-ESMF, the BHHLYP functional, still underestimates
the excitation energy by 2.74eV. Although there is no dipole change in this excitation,
the charge deformation in Rydberg states is still large since the virtual orbitals
are much more diffuse than and share little overlap with the occupied orbitals.
Consequently, the averaged $||\bm{X}||$ is as large as 0.25, 
comparable to that of CT excitations,
indicating that orbital relaxation is also important in Rydberg excitations. 

As discussed in the previous subsection, the main source of error in DFE-ESMF is 
the self-interaction error with approximate xc functionals:
The functional parametrized for the ground state is incapable of correcting the
self-interaction error of the excited electron residing in virtual orbitals. 
This problem is not too concerning in valence excitations since 
the occupied and virtual orbitals have similar characters, and hence similar amount 
of self-interaction. Therefore thanks to error cancellation, balanced description 
between the ground and the excited states can still be achieved. However, 
in Rydberg states, the virtual orbitals are much more diffuse than the occupied 
orbitals. Consequently, the amount of SIEs left in $J[n]$ after adding the UEG
exchange energy $E^{UEG}_x[n]$ to it, as in LDA, becomes clearly different in 
the occupied and virtual orbitals. This results in an unbalanced treatment 
between the ground and the excited state if one uses the same xc
functionals for both states, leading to a large error in this Rydberg excitation energy. 
In future, testing asymptotically corrected functionals in DFE-ESMF would appear to be warranted.

The unbalanced treatment of SIEs should not be exclusive 
to DFE-ESMF and could affect $\Delta$SCF DFT as well, since it also uses ground 
state functionals to describe excited states. In fact, in Figure \ref{fig:ne} we 
found that the size and trend of errors of $\Delta$SCF DFT with different xc 
functionals is quite comparable to DFE-ESMF, corroborating our speculations. 
In this case, the recovery of spin-symmetry in DFE-ESMF does not appear to have had much effect on the excitation energy.

\subsection{Multi-CSF Excited States}

We now turn our attention to states in which multiple CSFs are
important.
While DFE-ESMF can be formulated to treat such states and thus
may be expected to offer a significant advantage over
single-determinant methods such as $\Delta$SCF, such
optimism must be tempered by the increased risk of making
double counting errors and the complication of needing
a way to choose the values of the CI coefficients.
As we see in the following subsections, these issues
result in the present formulation of DFE-ESMF being
less effective in the multi-CSF case than it was in the
single-CSF cases discussed above.

\begin{figure}
\centering
\includegraphics[width=8.5cm,angle=0]{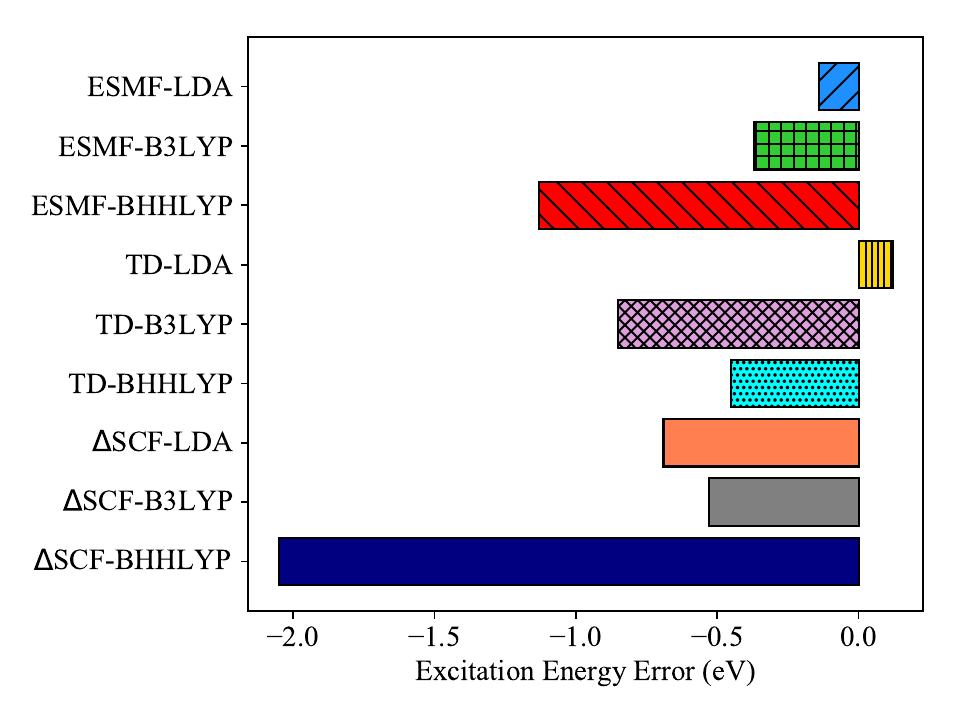}
\caption{The excitation energy error of singlet
         $\pi\rightarrow\pi^{\ast}$ excited states in CH$_2$O
         compared to EOM-CCSD results.
        }
\label{fig:ch2o_pi}
\end{figure}

\paragraph{CH$_2$O}
The $\pi\rightarrow\pi^{\ast}$
excited state in CH$_2$O contains significant contributions from two excited CSFs: 
HOMO-1$\rightarrow$LUMO and HOMO$\rightarrow$LUMO+2. 
In contrast to LiH, where TDDFT only predicts multiple
important CSFs for the BHHLYP functional,
this CH$_2$O excitation is strongly multi-CSF
in TDDFT regardless of the choice of funcitonal.
Excitation energy errors relative to
EOM-CCSD are plotted in Figure \ref{fig:ch2o_pi},
where the $\Delta$SCF results
have been obtained by optimizing the orbitals for the 
open-shell determinant that has the largest weight in
TDDFT's CI vector.
Curiously, both TDDFT and DFE-ESMF give the most accurate
prediction with LDA in this case, with DFE-ESMF showing
especially poor accuracy with BHHLYP.
This sharp contrast to the trends we saw in the single-CSF
cases is explained, at least in part, by the strong
dependence of the CI coefficients on the choice of
functional.
For LDA and B3LYP, the HOMO$\rightarrow$LUMO+2 CSF has a 
larger weight that the HOMO-1$\rightarrow$LUMO CSF,
whereas the relative importance of these CSFs
is reversed when using BHHLYP.
Noting that $\Delta$SCF places the energy of the
orbital-optimized HOMO-1$\rightarrow$LUMO CSF 1.86eV below
the energy of the orbital-optimized HOMO$\rightarrow$LUMO+2 CSF,
we can understand much of DFE-ESMF/BHHLYP's lowering of
the excitation energy simply in terms of the change
in the CI coefficients coming from TDDFT.
It therefore appears that, in future, re-optimizing the CI
coefficients within the DFE-ESMF framework may be important.

%Such a result is possibly due to 
%double counting of weak correlations of DFE-ESMF's multi-CSF formalism, which 
%can not be fully avoided as long as a multi-referenced wave function is used. 
%The $\Delta$SCF DFT results are obtained by optimizing the orbitals for the 
%open-shell determinant what has a larger weight in TDDFT's CI vector. We found 
%that it is less reliable than the other methods that preserve symmetry. Especially
%when paired with BHHLYP functional, choosing the determinant with larger weight from 
%two nearly equally contributing pieces produces more than 2eVs of error. 

%Due to the 
%charge transfer nature of this excited state, we found that the average of $||\bm{X}||$
%across different functionals is 0.31, larger than that of valence excitations. This is 
%consistent with our observations in LiH. 

\paragraph{CO}
\label{sec:co_hf}

\begin{figure}
\centering
\includegraphics[width=8.5cm,angle=0]{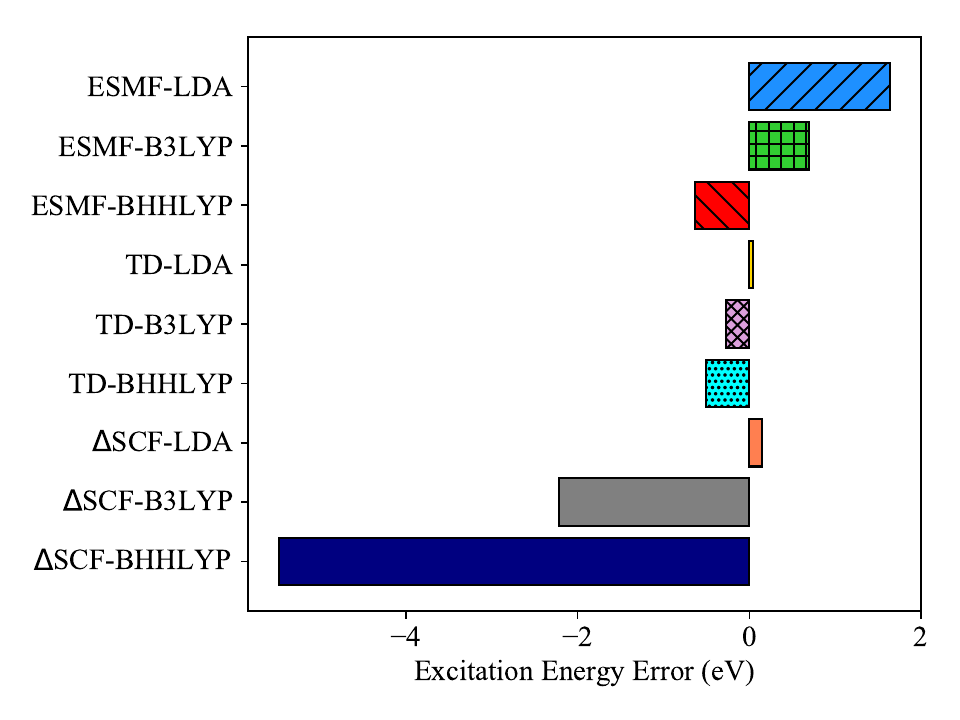}
\caption{The excitation energy errors relative to EOM-CCSD for the CO
         $\pi\rightarrow\pi^{\ast}$ singlet excitation. Energy minimization of 
         $\Delta$ SCF with BHHLYP functional ends up collapsing back to
         ground state. 
        }
\label{fig:co}
\end{figure}

In the $\pi\rightarrow\pi^{\ast}$ excitation in CO,
spatial symmetry ensures that the $p_x$ $\pi$
orbital is degenerate with its $p_y$ counterpart,
and likewise for the two $\pi^{\ast}$ orbitals, so that
the excited state is an equal mixture of the
$\pi_x\rightarrow\pi_x^{\ast}$ and
$\pi_y\rightarrow\pi_y^{\ast}$ CSFs.
The excitation energy errors for this state
are shown in Figure \ref{fig:co},  
where we have used our multi-CSF formalism to treat
this two-CSF state.
In this case, the DFE-ESMF trend is back in line with
what we saw in most other states, with BHHLYP providing
the most accurate result, although TDDFT is notably
more accurate than DFE-ESMF.
The trend for $\Delta$SCF is quite different, and it faces
multiple difficulties here, including the breaking of spin
and spatial symmetry as well as variational collapse
to the ground state with BHHLYP, which explains the unusually
large error in that case.

\paragraph{Ne}

\begin{figure}
\centering
\includegraphics[width=8.5cm,angle=0]{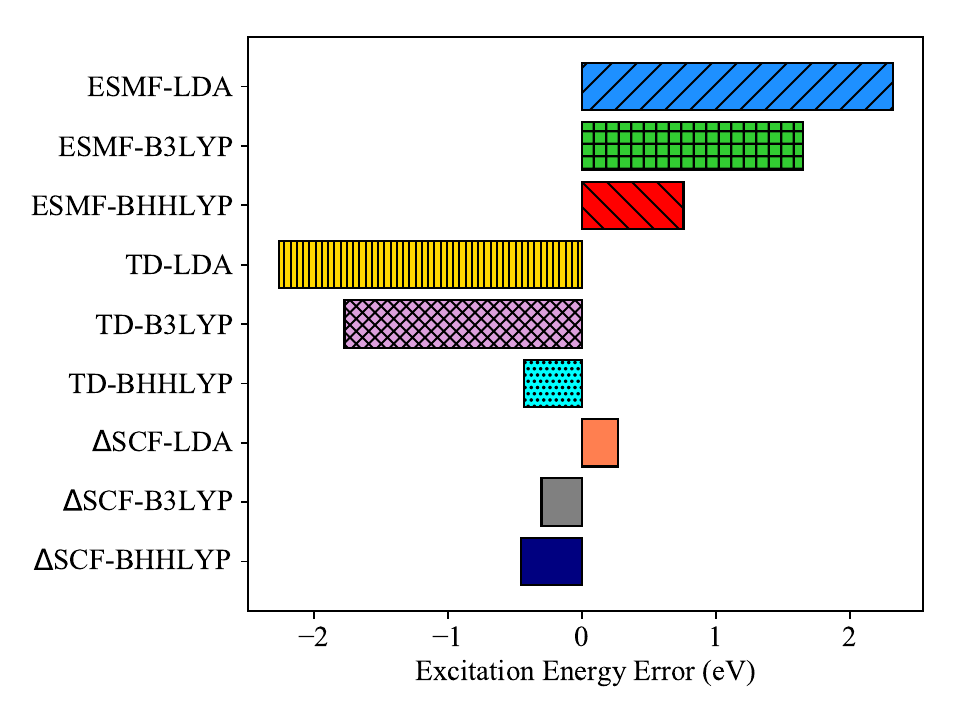}
\caption{Comparison of excitation energy errors relative to
         EOM-CCSD for the Ne 2p $\rightarrow$ 3p transition. 
        }
\label{fig:ne2}
\end{figure}

Unlike its 2s$\rightarrow$3s excitation, the Ne atom's 2p$\rightarrow$3p excited
state is a equal mixture of two CSFs.
In the plot of excitation energy errors (Figure \ref{fig:ne2})
we see that DFE-ESMF and TDDFT have similar accuracies in this case, although
the former errors high while the latter errors low.
Again, we see the tendency of DFE-ESMF to benefit from a relatively high fraction
of wave function exchange.
Unlike the other two multi-CSF cases discussed above, $\Delta$SCF proves
considerably more accurate for this excitation than either
DFE-ESMF or TDDFT.

%The averaged ||$\bm{X}$|| values across different functionals for the above three excited states 
%are 0.31, 0.24, and 0.18, respectively. This is as expected since the excited states 
%of CH$_2$O and CO are accompanied by small but non-negligible charge transfers, as 
%indicated by the dipole change, while the excited state of Ne has no charge transfer. 
%We also notice that the ||$\bm{X}$|| values in these multi-CSF valence excitations tend
%to be larger than those observed for
%in single-CSF valence excitations. This is as expected since excitations that involve 
%a larger number of orbitals should lead to larger amount of orbital relaxation effects. 

%\subsubsection{Comparison between Single- and Multi-CSF DFE-ESMF}

%Comparing single- and multi-CSF formalism of DFE-ESMF method, we find that the 
%latter is generally less accurate than the former. The main reason for this, 
%if we exclude the possibility of weak correlation double counting, is that the 
%multi-CSF formalism takes the CI coefficients of TDDFT and keeps them fixed. 
%As we see it

%This speculation is partly 
%confirmed by In Figure \ref{fig:ne} we also 
%show the error of $\Delta$SCF DFT, which we found
%to be in between that of TDDFT and DFE-ESMF. The broken of spin symmetry 

\section{Conclusions}
\label{sec:conclusion}

We have presented a density functional extension to the recently developed 
excited state mean-field theory in an attempt to
capture the effects of weak electron correlation.
%approach to excited states in DFT in which an approximated excited
%state variational principle is paired with a density functional energy expression.
By augmenting the ESMF wave function's energy expression with components
from density functional theory and inserting the resulting expression into
an approximate excited variational principle, 
the approach provides excited-state-specific orbital optimization in the
presence of a correlation treatment.
In the same way as KS-DFT closely parallels many aspects of HF theory,
this DFE-ESMF approach closely parallels ESMF theory.
Being a variational, time-independent approach, the method differs from
TDDFT in some important aspects, most notably in its ability to fully
relax orbitals and in the fact that it does not depend on the
ground state KS orbital eigenvalues and so avoids concerns related to
the EA/IP imbalance.
In preliminary testing, these advantages result in significantly improved
excitation energies for simple charge transfer examples, while
accuracy for other excitations is more comparable to TDDFT.
Compared to ROKS and $\Delta$SCF, which can also deliver full orbital
relaxation, DFE-ESMF is less prone to variational collapse thanks to its
use of an excited state variational principle.

Although these advantages and strengths are promising, the current
formulation of DFE-ESMF also has a number of shortcomings.
For excited states in which multiple CSFs make major contributions,
double counting becomes a serious concern and accuracy, although still
close to that of TDDFT, is reduced.
This challenge is especially concerning in the context of larger systems,
where multi-CSF states become increasingly common.
Rydberg states are also challenging, although this may have more to do
with the choice of density functional than with the DFE-ESMF approach
itself:  both it and TDDFT make substantial errors in these states.
Looking forward, we therefore see a number of directions for possible
improvement.
As we saw in all cases (except for the $\pi\rightarrow\pi^{*}$ transition
in formaldehyde where the sensitivity of TDDFT's CI coefficients to the
fraction of wave function exchange appears to be the culprit)
DFE-ESMF was most accurate when using a much higher fraction of wave
function exchange (50\%) than is typically found in ground state functionals.
This is not surprising given the open shell nature of excited states,
and suggests that a straightforward route to improved energetics may come
from retraining functionals in this direction.
Of course, this is just one question in a much broader array of functional design
issues, such as whether it would be advantageous to include range-separation or
asymptotic corrections, the latter of which may help with the Rydberg problem.
Functional design questions aside, our observation that accuracy tends to be
lower in multi-CSF states raises the question of whether this is due to
relying on TDDFT CI coefficients or comes from some other source.
If the former, then variational re-optimization of the CI coefficients
alongside the orbitals may be advantageous.
Finally, as a practical issue, our pilot implementation achieves the
desired cost scaling (equivalent to ground state KS-DFT)
but could be sped up considerably if implemented in a production level
quantum chemistry package.

\begin{acknowledgments}
This work was supported by the National Science Foundation's
CAREER program under Award Number 1848012.
Calculations were performed using the Berkeley Research
Computing Savio cluster. 
\end{acknowledgments}

\bibliography{DFE-ESMF}% Produces the bibliography via BibTeX.

\appendix
\section{Molecular Geometries}
\begin{figure}
\centering
\includegraphics[width=8.5cm,angle=0,scale=0.7]{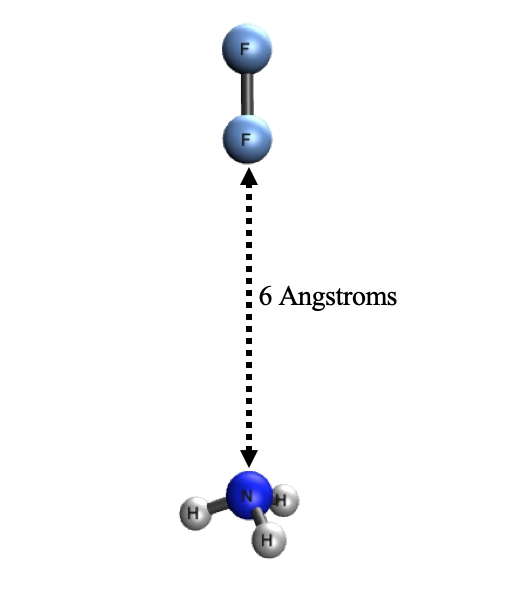}
\caption{Geometry of the NH$_3$-F$_2$ dimer. The N-H bond length and 
H-N-H bond angle is 1.02$\mathring{A}$ and 106.2$^\circ$. The F$_2$ bond length
is 1.43$\mathring{A}$. 
%{\color{red} Need to give FF and NH bond lengths.}
        }
\label{fig:nh3_f2_geo}
\end{figure}

In H$_2$O the H-O-H bond angle is chosen to be 104.5$^\circ$ and the O-H bond
length is 0.96$\mathring{A}$. In CH$_2$O the H-C-H bond angle is 116$^\circ$, 
and the C-H and C-O bond length are 1.11$\mathring{A}$ and 1.21$\mathring{A}$
respectively. In LiH the bond length is 1.6$\mathring{A}$, and the He-Be separation is 3.5$\mathring{A}$.
The geometry 
of the NH$_3$-F$_2$ dimer is shown in Figure \ref{fig:nh3_f2_geo}.

\section{Absolute Excitation Energies}
The absolute excitation energies studied in this work is shown in Table \ref{tab:esmf_eom_ccsd}, 
\ref{tab:tddft}, \ref{tab:roks}, and \ref{tab:dscf}. 

\begin{table}[h!]
\caption{Excitation energies of DF-ESMF and EOM-CCSD for the system studied in eV. Note that 
         the BHHLYP excitation energy of LiH $^3$($\sigma\rightarrow\sigma^{\ast}$) state is 
         from the single-CSF formalism, and the corresponding excitation energy is 3.07eV if the 
         multi-CSF formalism is used. 
         \label{tab:esmf_eom_ccsd}
}
\begin{tabular}{l r @{.} l r @{.} l r @{.} l r @{.} l r @{.} }
\hline\hline
  &
\multicolumn{2}{ c }{ \hspace{0mm} LDA \hspace{0mm} } &
\multicolumn{2}{ c }{ \hspace{0mm} B3LYP \hspace{0mm} } &
\multicolumn{2}{ c }{ \hspace{0mm} BHHLYP \hspace{0mm} } &
\multicolumn{2}{ c }{ \hspace{0mm} EOM-CCSD \hspace{0mm} } \\
\hline
H$_2$O $^1$(n$\rightarrow\pi^{\ast}$) & 9&07 & 8&66 & 8&26 & 8&07  \\
H$_2$O $^3$(n$\rightarrow\pi^{\ast}$) & 8&23 & 7&86 & 7&50 & 7&39 \\
LiH $^1$($\sigma\rightarrow\sigma^{\ast}$) & 4&62 & 4&23 & 3&60 & 3&47 \\
LiH $^3$($\sigma\rightarrow\sigma^{\ast}$) & 4&47 & 4&12 & 3&50 & 3&09  \\
CH$_2$O $^1$(n$\rightarrow\pi^{\ast}$) & 4&57 & 4&51 & 4&13 & 4&13 \\
CH$_2$O $^3$(n$\rightarrow\pi^{\ast}$) & 3&85 & 3&86 & 3&57 & 3&64 \\
CH$_2$O $^1$($\pi\rightarrow\pi^{\ast}$) & 9&94 & 9&71 & 8&93 & 10&08 \\
CO $^1$($\pi\rightarrow\pi^{\ast}$) & 7&13 & 6&17 & 4&86 & 5&49 \\
NH$_3$-F$_2$ & 10&59 & 10&12 & 9&03 & 9&29 \\
Ne 2s$\rightarrow$3s & 42&16 & 43&24 & 44&56 & 47&30\\
Ne 2p$\rightarrow$3p & 22&36 & 21&69 & 20&79 & 20&04\\
\hline\hline
\vspace{2mm}
\end{tabular}
\end{table}

\begin{table}
\caption{Excitation energies of TDDFT for the system studied in eV. 
         \label{tab:tddft}
}
\begin{tabular}{l r @{.} l r @{.} l r @{.} l r @{.} l r @{.} }
\hline\hline
  &
\multicolumn{2}{ c }{ \hspace{0mm} LDA \hspace{0mm} } &
\multicolumn{2}{ c }{ \hspace{0mm} B3LYP \hspace{0mm} } &
\multicolumn{2}{ c }{ \hspace{0mm} BHHLYP \hspace{0mm} } &
\multicolumn{2}{ c }{ \hspace{0mm} $\omega$B97X \hspace{0mm} } \\
\hline
H$_2$O $^1$(n$\rightarrow\pi^{\ast}$) & 7&34 & 7&54 & 8&14 & \multicolumn{2}{c}{\hspace{0mm} N/A \hspace{0mm}} \\
H$_2$O $^3$(n$\rightarrow\pi^{\ast}$) & 6&70 & 6&82 &  7&36 & \multicolumn{2}{c}{\hspace{0mm} N/A \hspace{0mm}} \\
LiH $^1$($\sigma\rightarrow\sigma^{\ast}$) & 3&16 & 3&34 & 3&65 & \multicolumn{2}{c}{\hspace{0mm} N/A \hspace{0mm}} \\
LiH $^3$($\sigma\rightarrow\sigma^{\ast}$) & 2&64 & 2&74 & 3&00 & \multicolumn{2}{c}{\hspace{0mm} N/A \hspace{0mm}} \\
CH$_2$O $^1$(n$\rightarrow\pi^{\ast}$) & 3&81 & 4&02 & 4&18 & \multicolumn{2}{c}{\hspace{0mm} N/A \hspace{0mm}} \\
CH$_2$O $^3$(n$\rightarrow\pi^{\ast}$) & 3&13 & 3&29 & 3&43 & \multicolumn{2}{c}{\hspace{0mm} N/A \hspace{0mm}} \\
CH$_2$O $^1$($\pi\rightarrow\pi^{\ast}$) & 9&23 & 9&63 & 10&12 & \multicolumn{2}{c}{\hspace{0mm} N/A \hspace{0mm}} \\
CO $^1$($\pi\rightarrow\pi^{\ast}$) & 5&52 & 5&21 & 4&99 & \multicolumn{2}{c}{\hspace{0mm} N/A \hspace{0mm}} \\
NH$_3$-F$_2$ & 0&00 & 1&88 & 5&32 & 5&84 \\
Ne 2s$\rightarrow$3s & 39&09 & 41&23 & 44&36 & 41&95 \\
Ne 2p$\rightarrow$3p & 17&78 & 18&27 & 19&60 & \multicolumn{2}{c}{\hspace{0mm} N/A \hspace{0mm}} \\
\hline\hline
\vspace{2mm}
\end{tabular}
\end{table}

\begin{table}
\caption{Excitation energies of ROKS for the system studied in eV. 
         \label{tab:roks}
}
\begin{tabular}{l r @{.} l r @{.} l r @{.} l r @{.} l r @{.} }
\hline\hline
  &
\multicolumn{2}{ c }{ \hspace{0mm} LDA \hspace{0mm} } &
\multicolumn{2}{ c }{ \hspace{0mm} B3LYP \hspace{0mm} } &
\multicolumn{2}{ c }{ \hspace{0mm} BHHLYP \hspace{0mm} } &
\multicolumn{2}{ c }{ \hspace{0mm} $\omega$B97X \hspace{0mm} } \\
\hline
H$_2$O $^1$(n$\rightarrow\pi^{\ast}$) & 8&15 & 6&75 & 7&79 & \multicolumn{2}{c}{\hspace{0mm} N/A \hspace{0mm}} \\
LiH $^1$($\sigma\rightarrow\sigma^{\ast}$) & 3&48 & 3&41 & 3&37 & \multicolumn{2}{c}{\hspace{0mm} N/A \hspace{0mm}} \\
CH$_2$O $^1$(n$\rightarrow\pi^{\ast}$) & 3&94 & 3&79 & 3&68 & \multicolumn{2}{c}{\hspace{0mm} N/A \hspace{0mm}} \\
NH$_3$-F$_2$ & -5&50 & -5&53 & -5&26 & -5&24 \\
\hline\hline
\vspace{2mm}
\end{tabular}
\end{table}

\begin{table}
\caption{Excitation energies of $\Delta$SCF-DFT for the system studied in eV. 
         \label{tab:dscf}
}
\begin{tabular}{l r @{.} l r @{.} l r @{.} l r @{.} l r @{.} }
\hline\hline
  &
\multicolumn{2}{ c }{ \hspace{0mm} LDA \hspace{0mm} } &
\multicolumn{2}{ c }{ \hspace{0mm} B3LYP \hspace{0mm} } &
\multicolumn{2}{ c }{ \hspace{0mm} BHHLYP \hspace{0mm} } &
\multicolumn{2}{ c }{ \hspace{0mm} $\omega$B97X \hspace{0mm} } \\
\hline
CH$_2$O $^1$($\pi\rightarrow\pi^{\ast}$) & 9&38 & 9&55 & 8&03 & \multicolumn{2}{c}{\hspace{0mm} N/A \hspace{0mm}} \\
CO $^1$($\pi\rightarrow\pi^{\ast}$) & 5&64 & 3&26 & 0&00 & \multicolumn{2}{c}{\hspace{0mm} N/A \hspace{0mm}} \\
NH$_3$-F$_2$ & 9&45 & 9&06 & 8&38 & 9&10 \\
Ne 2s$\rightarrow$3s & 41&18 & 42&57 & 44&36 & 43&15 \\
Ne 2p$\rightarrow$3p & 20&31 & 19&73 & 19&58 & \multicolumn{2}{c}{\hspace{0mm} N/A \hspace{0mm}} \\
\hline\hline
\vspace{2mm}
\end{tabular}
\end{table}

\end{document}